\documentclass[preprintnumbers,nofootinbib,prd,12pt]{revtex4}

\usepackage{graphicx}
\usepackage{wasysym}
\usepackage{longtable}
\usepackage[hypertex]{hyperref}

\newcommand{\lsim}{\lesssim}

\newcommand{\beq}{\begin{equation}}
\newcommand{\eeq}{\end{equation}}

\newcommand{\unit}[1]{\,\mathrm{#1}}

\begin{document}

\pagestyle{plain}

\preprint{BNL-HET-08/9}
\preprint{CERN-PH-TH-2008-081}
\preprint{VPI-IPNAS-08-09}
\title{Feasibility Study for Measuring Geomagnetic Conversion of Solar
Axions to X-rays in Low Earth Orbits}

\author{Hooman Davoudiasl}
\email{hooman@bnl.gov}

\affiliation{Department of Physics, Brookhaven National Laboratory,
  Upton, NY 11973, USA}

\author{Patrick Huber}
\email{pahuber@vt.edu}

\affiliation{Physics Department, Theory Division, CERN,
1211 Geneva 23, Switzerland}

\affiliation{Department of Physics, Virginia Tech, Blacksburg, VA 24062, USA}


\begin{abstract}
  We present a detailed computation of the expected rate for
  Geomagnetic Conversion of Solar Axions to X-rays (GECOSAX) along the
  orbit of an x-ray satellite. We use realistic satellite orbits and
  propagation in time. A realistic model for the Earth's magnetic
  field, which properly accounts for its spatial non-uniformity, is
  used. We also account for the effect of the Earth's atmosphere on
  the propagation of x-rays in our calculation of axion-photon
  conversion probability.  To estimate possible sensitivities to the
  axion-photon coupling $g_{a\gamma}$, we use an actual measurement of
  the expected backgrounds by the SUZAKU satellite.  Assuming a
  detector area of $10^3\,\mathrm{cm}^2$ and about $10^6\,\mathrm{s}$
  of data, we show that a $2\,\sigma$ limit of $g_{a\gamma} <
  (4.7-6.6)\times 10^{-11}\,\mathrm{GeV}^{-1}$ from GECOSAX is
  achievable, for axion masses $m_a<10^{-4}\,\mathrm{eV}$.  This
  significantly exceeds current laboratory sensitivities to
  $g_{a\gamma}$.

\end{abstract}

\date{\today}

\maketitle

\section{Introduction}

Weakly interacting light pseudo-scalars are well-motivated in particle
physics.  For example, experimental observations require the size of
$CP$ violation in strong interactions, parametrized by the angle
$\theta$, to be quite small: $\theta \lsim 10^{-10}$.  However, the
symmetries of the Standard Model (SM) allow $\theta\sim 1$; this is
the strong $CP$ problem.  An elegant solution to this puzzle was
proposed by Peccei and Quinn (PQ) \cite{Peccei:1977hh,Peccei:1977ur},
where a new $U(1)$ symmetry, anomalous under strong interactions, was
proposed.  This $U(1)$ symmetry is assumed to be spontaneously broken
at a scale $f_a$, resulting in a pseudo-scalar Goldstone boson $a$
\cite{Weinberg:1977ma,Wilczek:1977pj}, the axion.  Non-perturbative
QCD interactions at the scale $\Lambda_{\rm QCD} \sim 100\unit{MeV}$
generate a potential and hence a mass $m_a^{PQ} \sim \Lambda_{\rm
  QCD}^2/f_a$ for the axion.  Experimental and observational bounds
have pushed $f_a$ to scales of order $10^7\unit{GeV}$ or more.  As
$f_a$ sets the inverse coupling of the axion to the SM fields, the
current data suggests that axions are basically `invisible' and very
light.  Here we note that some cosmological considerations related to
the overclosure of the universe suggest an upper bound $f_a \lsim
10^{12}$~GeV for the PQ
axion~\cite{Preskill:1982cy,Abbott:1982af,Dine:1982ah,Turner:1985si}.
Apart from the considerations related to the strong $CP$ problem,
axion-like particles are ubiquitous in string theory.  In addition,
axion-like particles have been used in various astrophysical and 
cosmological 
models\footnote{See, for 
example, Ref.~\cite{Csaki:2001yk}.}.  
In the following, the term axion is generically used to refer to any of the
above, or other, possible instances of such weakly interacting light
pseudo-scalars.

The coupling of the axion to photons is given by \cite{GGR}
\beq
{\cal
L}_{a\gamma} = -\frac{a}{4 M} F_{\mu \nu} {\tilde F}^{\mu\nu}
= g_{a\gamma}\, a\, \vec{E}\cdot \vec{B},
\label{La}
\eeq where $M \sim (\pi/\alpha)f_a$ and $\alpha \simeq 1/137$ is the
fine structure constant.  $F_{\mu\nu}$ is the electromagnetic field
strength tensor, ${\tilde F}^{\mu\nu}$ is its dual, $g_{a\gamma}
\equiv M^{-1}$ is the axion-photon coupling; $\vec{E}$ and $\vec{B}$
are the electric and magnetic fields, respectively, corresponding to
$F_{\mu\nu}$.  The interaction in (\ref{La}) makes it possible for hot
plasmas, like the Sun, to emit a flux of axions through the Primakoff
process \cite{Pirmakoff:1951pj}.  This same interaction has also led
to experimental proposals \cite{Sikivie:1983ip} for detecting the
axion through its conversion to photons in external magnetic fields.
Various experimental bounds, most recent of which is set by the CAST
experiment \cite{Andriamonje:2007ew}, suggest that $g_{a\gamma} \lsim
10^{-10}\unit{GeV}^{-1}$.  For a review of different bounds on axion
couplings, see Ref.~\cite{Yao:2006px}.

In what follows, we study the feasibility of a recently proposed
approach for detecting solar axions with an x-ray telescope in orbit
\cite{Davoudiasl:2005nh}, based on geomagnetic conversion of solar
axions to x-rays (GECOSAX)\footnote{ The possibility of using
  planetary magnetic fields as a conversion region for high energy
  cosmic axions was discussed in Ref.~\cite{Zioutas:1998ra}.}.  The
estimate of the expected x-ray flux presented in
Ref.~\cite{Davoudiasl:2005nh} was based on a number of simplifying
assumptions:
\begin{enumerate}
\item The satellite orbit was  a circle.
\item The orbit was aligned to lie in the equatorial plane of the
  Earth.
\item The Earth axis was perpendicular to the Ecliptic.
\item The available conversion length was taken to be the altitude of
  the satellite.
\item The magnetic field was assumed uniform and perpendicular to the
  direction of axion propagation.
\item The effect of the Earth atmosphere was neglected.  Consequently,
the effective mass of the photon in medium was ignored; $m_\gamma\to 0$.
\end{enumerate}
These assumptions allowed a treatment of GECOSAX within the same
formalism relevant for helioscopes \cite{Sikivie:1983ip,vanBibber:1988ge},
{\it i.e.} axion-photon conversion
{\it in vacuo} in a constant magnetic field which is perpendicular to
the direction of the axion momentum. The conversion rate , in the
limit of vanishing axion mass $m_a \to 0$, is then
simply given by
\begin{equation}
P_{a\gamma}^s = 2.45\times10^{-21}\,
\left(\frac{g_{a\gamma}}{10^{-10}\,\mathrm{GeV}^{-1}}\right)^2\,
\left(\frac{B}{\mathrm{T}}\right)^2\,\left(\frac{L}{\mathrm{m}}\right)^2\,.
\label{eq:simple}
\end{equation}
Taking $g_{a\gamma}=10^{-10}\,\mathrm{GeV}^{-1}$ and using CAST
parameters $B=9\,\mathrm{T}$ and $L=10\,\mathrm{m}$, we obtain
$P_{a\gamma}^s\simeq2\times10^{-17}$. Replacing the CAST magnet with the
geomagnetic field and taking $L$ to characterize a low-Earth-orbit,
we have $B=3\times10^{-5}\,\mathrm{T}$ and
$L=6\times10^5\,\mathrm{m}$ and thus get for
$P^s_{a\gamma}\simeq 8\times10^{-19}$, which is only a factor of about
25 smaller than the CAST conversion probability.  However, in
Ref.~\cite{Davoudiasl:2005nh}, it was noted that this can
typically be overcompensated
by the larger detection area of an orbiting x-ray
telescope. In~\cite{Davoudiasl:2005nh} it was shown that the resulting
x-ray signal on the dark side of the Earth would have a number of
unique features which would make it very hard to be mistaken for
anything else: upward going x-rays, $T=4\,\mathrm{keV}$ black
body spectrum, direction from within $3'$ from the center of the Sun
and characteristic modulation with $B^2L^2$.

In the following, we would like to address the simplifying assumptions
one by one and illustrate their effect on the actual axion-photon
conversion rate. In section~\ref{sec:orbit}, we will discuss how to
account for the proper satellite-Sun-Earth geometry.
We will also discuss how to compute the actual position of any
given satellite. This will address assumptions 1-4. Next we will
address assumption~5 in section~\ref{sec:geomagnet}
with a proper magnetic model. In section~\ref{sec:propagation}, we
will present a full treatment of axion-photon conversion in a
dispersive and absorptive medium, using a model of the Earth's
atmosphere, which addresses assumption~6. In
section~\ref{sec:fluxes}, we will present the resulting x-ray fluxes
for various satellites, followed by a discussion of achievable
sensitivities for $g_{a\gamma}$, in section~\ref{sec:sens}.
Finally, we will present a discussion of our results and
the future outlook, in section~\ref{sec:conc}.

\section{Geometry and satellite orbits}
\label{sec:orbit}

The basis for any detailed calculation of the axion-photon conversion
rate is a correct description of the geometry. We introduce the following
notation: plain capital Latin or Greek letters denote a point, where
$O$ denotes the origin of our coordinate system. If $P$ is a point, then
its position vector $\overline{OP}$ is denoted by $\vec{p}$. For any
vector $\vec{v}$ its length is denoted by $v$. The unit vector along
the direction of $\vec{v}$ is denoted by $\vec{e}_v=\vec{v}/v$. The Sun
is at $S$, the center of the Earth is at $E$ and the x-ray satellite
is at $X$. For any point $P$, we define its height vector $\vec{h}_P$
as the vector which goes from the surface of the Earth to $P$ and is
perpendicular to the tangent plane of the Earth surface at its
starting point. The starting point is called the footprint of $P$ and
denoted by $P_F$. This definition may seem involved, however, it also
holds for the actual geoid and not only for a spherical Earth.

The abstract definition of the problem can be easily done without
specifying a coordinate system. The actual numerical calculation, of
course, has to specify a definite coordinate system, which is relegated
to appendix~\ref{app:coords}.

\begin{figure}
\includegraphics[width=\textwidth]{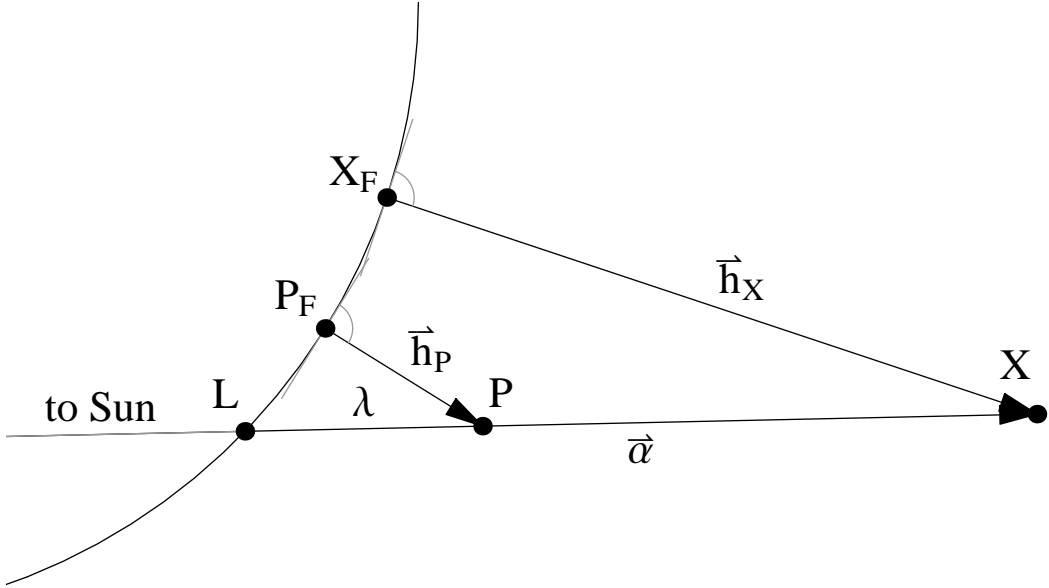}
\caption{\label{fig:geo} Geometry of the GECOSAX configuration drawn
  in the plane spanned by the center of the Earth, the center of the
  Sun and the satellite's position at $t_0$.}
\end{figure}

In Fig.~\ref{fig:geo} a two dimensional schematic view of the problem
is shown at one instant in time $t_0$. Since axions/photons travel
close to/at the speed of light $c$, they experience an essentially
static environment. The geometry only changes notably on timescales
large compared to the propagation time $\tau_p \sim
10^6\,\mathrm{m}/c\sim10^{-3}\,\mathrm{s}$. Therefore, the geometry
can be regarded as fixed at any given time $t_0$. The first task is to
determine the path traveled by the axions. The axions propagate in a
straight line from the Sun $S$ to the satellite $X$.  For axion
conversion, however, only the part of $\overline{SX}$ which is on the
dark side of the Earth is relevant. Thus, the intersections of
$\overline{SX}$ with the surface of the Earth have to be found. We
account for the ellipsoidal shape of the Earth and use the so called
WGS72 parameters\footnote{For polar orbits the difference in radius of
  a spherical and ellipsoidal Earth can cause up to a $10\,\mathrm{s}$
  difference in the duration of the dark orbit.}. In general there can
be none, one or two such intersections.  The solution on the dark side
will be denoted by $L$. This allows us to define the line of sight
(LOS) $\vec{\alpha}=\overline{LX}$. Note, that this definition of
being on the dark side is purely geometrical and neglects the angular
diameter of the Sun, atmospheric refraction and absorption. The
angular diameter of the Sun reduces the useful part of the orbit by
about $4\,\mathrm{s}$. This follows from the fact that the Sun
subtends $0.5^\circ$ and the satellite travels a full circle in about
90 minutes. The effect of the Earth atmosphere is quite a bit larger
since it becomes non-transparent for x-rays below an altitude of about
$50-100\,\mathrm{km}$, thus increasing the effective radius by that
amount, which by explicit calculation would increase the dark orbit
duration by a few times $10\,\mathrm{s}$. This overcompensates for our
neglect of the solar diameter and makes our overall treatment
conservative. Using the position of $L$ as defined above, one can
parametrize the position of any point along the line of sight

\begin{equation}
\label{eq:los}
\vec{p}(\lambda)\equiv\lambda\,\vec{e}_\alpha+\vec{l}\,.
\end{equation}

In the course of the actual calculation the height $h_P$ of $P$ is
needed, since the air density is a function of the actual height. Note that
the height of the satellite $h_X$ is always smaller than the length of
the LOS, {\it i.e.} $h_X\leq \alpha $. This implies that assumption~4
is in fact conservative, and we will find that the relative increase
of $\alpha$ with respect to $h_X$ will compensate largely for the
losses in x-ray flux due to the other effects considered in the
following. The algorithm for the solar position is taken
from~\cite{Meeus} and it's accuracy is better than $1^\prime$.

The idealized orbit of any satellite is a solution to the Kepler
problem. Thus, knowing the satellite's position and velocity at time
$t_0$, it is possible to predict its future position at $t_1$. In
reality, there are, however, various factors which lead to deviations
from the simple Kepler orbit, among which are: non-vanishing higher
multipoles of the mass distribution of the Earth, atmospheric
drag, gravitational influence from the Moon (and to a lesser degree
from the Sun), {\it etc.}. The prediction of satellite orbits is, of
course, a matter of great importance for operators of satellites and
is also needed for military applications. The aforementioned effects
disturbing the simple Kepler orbit can be accounted for in a
general\footnote{General in the sense, that the resulting theory is
  applicable to a wide, general class of orbits and not restricted to
  particular orbits like {\it e.g.} ones with a low eccentricity.}
perturbation theory. Many of the perturbations are well know and can
be quite exactly computed. From the observation of the actual position
and velocity of a satellite at time $t_0$ it is possible to extract
the unperturbed Kepler orbit, which would follow in the absence of any
perturbing factors. In order to obtain an accurate prediction for the
future, a specific set of perturbations is taken into account. In
doing this, it is crucial that the initial unperturbed Keplerian orbit
data is extracted using a model which is compatible with the algorithm
used for future positions.

One standard format is the so called `NORAD\footnote{NORAD is the
  North American Aerospace Defense Command.} element sets', and a
description of the perturbation model called SGP4\footnote{SGP4 stands
  for `simplified general perturbation version 4'. Historically, one
  distinguished SGP4 and SDP4, where the latter one is used for
  `deep-space' orbits with periods longer than 225 minutes. For most
  parts of this work, we use only SGP4, {\it i.e.} orbits with periods
  smaller than 225 minutes.} can be found in~\cite{spacetrack3}. Since
satellite propagation is done in perturbation theory, errors
inevitably will accumulate and render the predictions unreliable.
Therefore, element sets for basically all active satellites are issued
periodically by NORAD and made accessible at~\cite{tle} in the so
called `two line element' (TLE) format. The implementation of the
NORAD orbit prediction algorithm we use is taken from the {\tt
  predict} program, which is an open-source C language satellite
tracking software~\cite{predict}\footnote{In reality, it seems that
  all implementations found in open accessible sources go back
  to various, different original implementations by
  T.S~Kelso~\cite{tle} in a number of programming languages. The one
  we are using is no exception.}. It directly takes the TLE of a
satellite, a time $t_0$ and returns its position in the
ECI\footnote{This is an Earth Centered Inertial set of coordinates,
  discussed in appendix~\ref{app:coords}.} at $t_0$. All satellites
are indexed by NORAD using so called US SPACECOM identification
numbers, these are 5 digit numbers starting with 00001 for the SPUTNIK
satellite. We will use these 5 digit numbers to refer to all
satellites in this paper, any satellite names are written in capitals.
A list if all US SPACECOM IDs and the corresponding names is given in
table~\ref{tab:tle}.

Some remarks about our use of TLEs and SGP4 are in order. We use SGP4
since it is the simplest general purpose algorithm and the necessary
input data, the TLEs, are easily available. SGP4 is by no means state
of the art, it was developed to allow reliable tracking of thousands
of objects with the limited computing power available in the 60s and
70s.  Clearly, in an actual experiment one would use telemetry data
and direct numerical integration, possibly even GPS, thus reducing any
position errors to around $100\,\mathrm{m}$ or less~\cite{tleerror}.
The accuracy of predictions made with SGP4 relative to GPS position
determination was studied in detail in~\cite{tleerror}. The typical
errors for SGP4 are about $\pm5\,\mathrm{km}$ cross-track, {\it i.e}
perpendicular to the satellite momentum, within $\pm 15$ days from the
epoch of the used TLE, whereas the in-track error, {\it i.e.} along
the orbit, is about $\pm 20\,\mathrm{km}$. For a satellite moving at
about $10\,\mathrm{km}/\mathrm{s}$ this gives rise to a timing error
of about $2\,\mathrm{s}$. We checked that these errors have, in fact,
a very small impact on the average axion signal, since the satellite
still reaches every point under more or less the same circumstance
with respect to magnetic field orientation and direction to the Sun.
For this test, we used historic TLEs of satellite 27370 (RHESSI),
obtained from~\cite{tle}, issued about 8 months apart. We found that
there was a time difference of several minutes in when the satellite
entered the dark orbit, but once we corrected for this time shift, the
GECOSAX fluxes where identical to within $\sim10\%$.
In~\cite{tleerror} it is shown that the inter-TLE variation is a good
indicator for the actual accuracy. This is clearly an extreme example
since TLEs are re-issued about every other week.  Thus, we conclude
that SGP4 with current TLEs is accurate enough by a large margin.

To summarize, we found that the corrections due to the proper
treatment of the geometry are quite large and lead to a pronounced
variation of the expected flux along each orbit because of changes in
the length $\alpha$ of the LOS. Geometry related effects are accounted
within about $\pm25\,\mathrm{km}$ or $\pm 5\,\mathrm{s}$ in our
calculation, which introduces less than $10\%$ error in the GECOSAX
signal prediction. The errors introduced by our simplified treatment of
coordinate transformation in appendix~\ref{app:coords} are negligible
in comparison to the intrinsic errors of the satellite orbit prediction.

\section{Magnetic field of the earth}
\label{sec:geomagnet}

To a first approximation, the magnetic field of the Earth is a dipole
whose axis intersects the surface at the magnetic poles, which do not
coincide with the geographic ones. This mis-alignment of rotation and
magnetic axes alone would induce a typical periodic variation of the
x-ray flux produced by GECOSAX. However, the geomagnetic field has
various other irregularities and deviations from a simple dipole form.

\begin{figure}[t!]
\includegraphics[width=\textwidth]{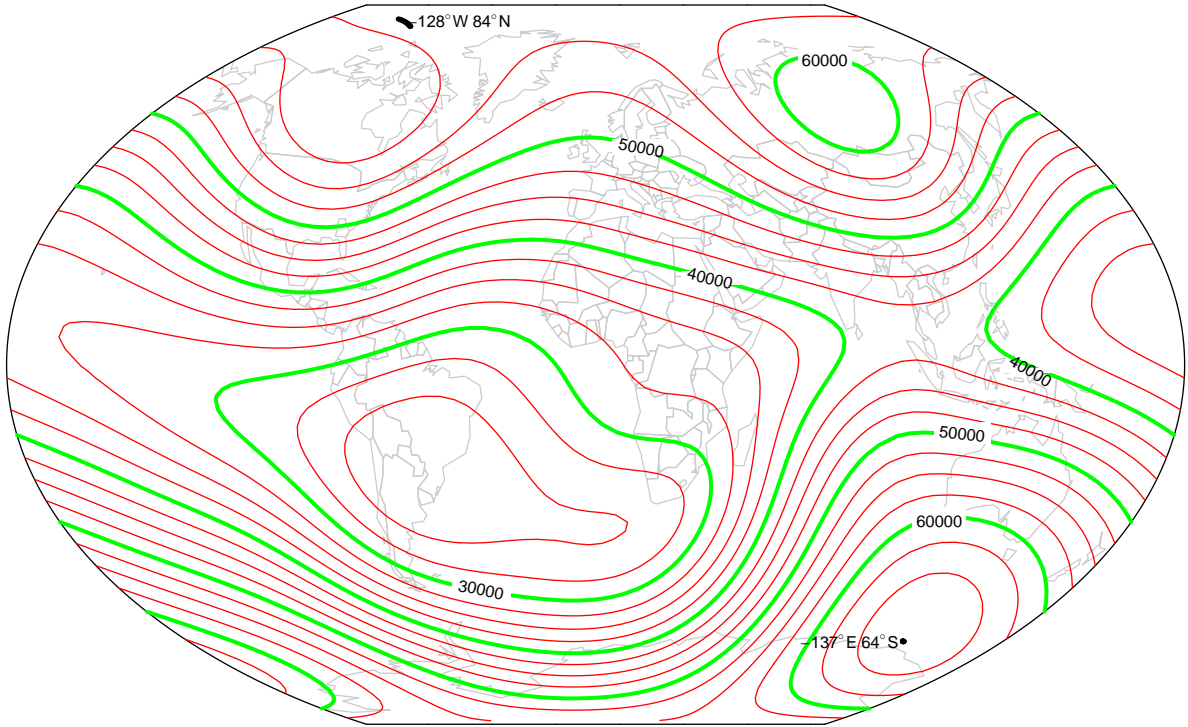}
\caption{\label{fig:bfield} Map of the total magnetic field strength
  at sea level for 2008.5~\cite{wmm}. Red, thick contours are in
  steps of $10\,000\,\mathrm{nT}$, thin red ones are in steps of
  $2\,500\,\mathrm{nT}$. The number on top of the red, thick contours
  are the magnetic field strength in nT. The black dots denote the
  positions of the magnetic dip pole for each year from 2005 till
  2010. The coordinates give the position of the dip pole in 2008.5.
  The map is a Winkel tripel projection.}
\end{figure}

We use a realistic 3-d model of the magnetic field of the earth, the
so called World Magnetic Model 2005~\cite{wmm1}, which is available in
machine readable form at~\cite{wmm}. The World Magnetic Model is the
standard model of the US Department of Defense, the UK Ministry of
Defense, the North Atlantic Treaty Organization (NATO), and the World
Hydrographic Office (WHO) for navigation and attitude/heading
referencing systems.  It is intended to meet its stringent error
specifications (better than $1\%$) from sea level up to an altitude of
$600\,\mathrm{km}$. Since it is given as a series expansion in
spherical harmonics, mathematically it stays well-defined out to
larger radii.  If there were no electrical currents in the upper
atmosphere and no solar wind, {\it i.e.}  additional sources of
magnetic fields outside the Earth, the model would be accurate up to
many Earth radii.  In practice, interactions with the solar wind and
atmospheric electrical currents produce magnetic fields of around
$100-500\,\mathrm{nT}$, at an altitude of $\sim 1000\,\mathrm{km}$,
during magnetically quiet times and perturbations can reach up
$2000\,\mathrm{nT}$ during strong magnetic storms. Magnetic
perturbations are indexed by the $A_p$ index, which is the daily
average of the $a_p$ index. It denotes the deviation from the most
disturbed component of the local magnetic field vector from its mean,
undisturbed value in units of $2\,\mathrm{nT}$. The $A_p$ index is
derived from the observations of 11 geomagnetic observatories and has
been regularly collected since 1932. $A_p$ values larger than 100 are
classified as indicating a severe magnetic storm. Only 1\% of the days
from 1932 till 1992 have reported a value of $A_p>100$~\cite{coffey}.
Hence, for almost all observation conditions the errors introduced by
the day to day variability of the geomagnetic field will be small.
Thus, the errors introduced by using the magnetic model up to an
altitude of $1000\,\mathrm{km}$ are certainly less than 10\%, most
likely much less than $5\%$~\cite{handbuchderphysik,maus,ecss}.
Therefore, in principle, it seems feasible to extend the permissible
range of altitudes maybe up to 1-2 Earth radii, however, this would
require a more careful analysis of the external magnetic fields, which
is beyond the scope of this work. We, therefore, will restrict all
analysis to altitudes below $1\,000\,\mathrm{km}$, unless otherwise
mentioned.

The magnetic model also includes a prediction of the annual
variation of geomagnetic parameters from 2005 to 2010.  From these
variations we expect a less than 1\% annual change in the relevant
parameters. Given this 3-d vectorial map, we compute the transverse
$\vec{B}$-field along the axion path. The total field strength is
shown in Fig.~\ref{fig:bfield}.

\section{Axion propagation}
\label{sec:propagation}

The axions and x-rays will have to traverse the upper Earth
atmosphere, which causes absorption and refraction of x-rays and hence
will also influence the axion-photon conversion probability.  To a
rough approximation the interaction of x-rays with an energy of few
keV with air can be described by Thomson scattering from free
electrons~\cite{henke}. Air mostly consists of nitrogen and oxygen
having atomic numbers $Z$ of 7 and 8, respectively.  The binding
energies of the innermost electrons thus are about $Z^2\,13.6
\,\mathrm{eV}\simeq 600-800\,\mathrm{eV}$ and thus small compared to
the photon energy for most of the range of interest. On the other
hand, the photon energies are very small compared to the rest mass of
the electron and hence the scattering is highly non-relativistic and
pair creation cannot take place. Thus, all effects on x-ray
propagation should be a function of the electron density, which itself
closely traces the mass density of air.  At standard temperature and
pressure (STP) of $273.15\,\mathrm{K}$ and $101\,325\,\mathrm{Pa}$, we
use as volume (molar) fractions $78.1\%$ N$_2$, $21.0\%$ O$_2$, and
$0.93\%$ Ar\footnote{With $Z=18$, Ar has binding energies in the keV
  range.  However, due to its small molar fraction we may ignore it
  for the exposition, here. In the numerical analysis it is accounted for.}.

The absorption length $\lambda = \Gamma^{-1}$ for x-rays of energy
$1-10\,\mathrm{keV}$, in air at STP, has been obtained
from~\cite{henke,xraydata}. We have assumed that x-ray absorption
scales with the electron density along the axion path. Assuming a
constant composition of the atmosphere with altitude, the electron
density\footnote{We will comment on this in more detail, later in this
section.} is directly proportional to the mass density.

To include refraction, we use the effective photon mass $m_\gamma$
given by~\cite{vanBibber:1988ge}
\beq
m_\gamma^2 = 4 \pi r_0 [\rho f_1/(A m_u)] ,
\label{mgam}
\eeq where $r_0\simeq 2.82\times 10^{-15}$~m is the classical electron
radius, $A$ is the atomic mass number of the gas (atmosphere), $m_u$
is the atomic mass unit, $\rho$ is the gas density, $f_1 \simeq Z$,
and $Z$ is atomic number of the gas.  This formula can be generalized
for a compound gas, like the air, by noting that the quantity $\rho
Z/(A m_u)$ is the electron number density $n_e$ for the medium, which
in the above equation is assumed to be made up of only one element.

For a simplified derivation we assume that air is composed of $78\%$
N$_2$ and $22\%$ O$_2$, by volume. In the ideal gas limit, the volume
and molar fractions are the same.  Given that for dry air the molar
density is $\rho_{air} \simeq 44.48$~mol m$^{-3}$ (STP), we find that
$n_e \simeq 44.48\, N_A (0.78\times 14 + 0.22\times 16) \simeq 3.87
\times 10^{26}$~m$^{-3}$, where $N_A$ is Avogadro's number.  Then,
Eq.~(\ref{mgam}) yields
\beq
m_\gamma=0.64\left(\frac{\rho}{\mathrm{kg}\,\mathrm{m}^{-3}}\right)^{1/2}
\mathrm{eV}\,.
\label{eq:mgam_air}
\eeq
We checked that using the full energy dependence of $f_1$ given
in~\cite{henke,xraydata} and a more detailed air composition does not
change $m_\gamma$ by more than $2\%$.

As the following general argument will show, the axion conversion path
length is only logarithmically sensitive to density variations. In all
density models, the density profile can be locally described by an
exponential with an altitude dependent scale height $H(h)$
\begin{equation}
\label{eq:expdens}
\rho(h)\equiv \rho_0 e^{-\frac{h}{H(h)}}\,,
\end{equation}
where both $\rho_0$ and $H(h)$ can be time dependent.  In order to
estimate the impact on axion propagation, we need to understand up to
what altitude absorption plays a role and what impact a finite photon
mass has. In order to asses the sensitivity to changes in absorption
we can compute the escape probability $p_{\mathrm{esc}}$ of a x-ray
photon from a given altitude $\eta$ to infinity:
\beq
p_\mathrm{esc}=e^{-c_p}\quad\mathrm{with}\quad
c_p= \int_{\eta}^{\infty}dx\,\Gamma(x)\,,
\eeq
where $\Gamma(x)$ is
the inverse absorption length and $c_p$ the so called absorption
coefficient. $\Gamma(x)$ itself is a function of the density and
is given by
\beq \Gamma(x)\equiv \mu \rho(x)\,,
\eeq
where $\mu$ is the mass
attenuation coefficient. We now can define the escape altitude
$a_\mathrm{esc}$ by demanding $p_\mathrm{esc}=1/e$. This then
translates to the following condition
\beq
1\stackrel{!}{=}c_p=\int_{\eta_\mathrm{esc}}^{\infty}dx\,
\Gamma(x)=\int_{\eta_\mathrm{esc}}^{\infty}dx\,\mu\,\rho_0
e^{-\frac{x}{H(x)}}\,.
\eeq
Assuming that $H(x)=H=const.$ we can easily solve for $x$ and obtain
\beq
\eta_\mathrm{esc}=H\ln (H \rho_0\, \mu)\,.
\eeq

The effect of a non-vanishing $m_\gamma$ is the same as the one of a
non-vanishing axion mass: they both affect the oscillation or
coherence length for photon-axion conversion. The conversion
probability will be suppressed whenever the oscillation length is
short compared to the available path length. For the following
discussion we define the oscillation length $l_\gamma$
to be due to the finite photon mass
\beq l_\gamma=\frac{4\pi \omega}{m_\gamma^2}\,,
\eeq
where $\omega$ denotes the photon energy. In order for $m_\gamma$ to be
negligible, we require $l_\gamma<L\simeq1000\,\mathrm{km}$. Since
$m_\gamma\propto\sqrt{\rho}$, it follows that
$l_\gamma\propto\rho^{-1}$. The altitude $\eta$ at which $l_\gamma$
reaches a certain value can be computed, using
Eqs.~\ref{eq:mgam_air} and~\ref{eq:expdens}
\beq
\eta=H\ln \left[1.65\times 10^5
  \,\left(\frac{l_\gamma}{\mathrm{km}}\right)\,\left(\frac{\rho_0}{\mathrm{kg}\,\mathrm{m}^{-3}}\right)\,\left(\frac{\mathrm{keV}}{\omega}\right)\right]\,.
\eeq
Let $\eta_\gamma$ denote the value of $\eta$ at $l_\gamma=1000\,\mathrm{km}$.
Again we find that the dependence of $\eta_\gamma$ on
$\rho_0$ is only logarithmic. Therefore, all atmospheric effects
will depend only weakly on the precise value of density.

\begin{figure}[t!]
\includegraphics[width=\textwidth]{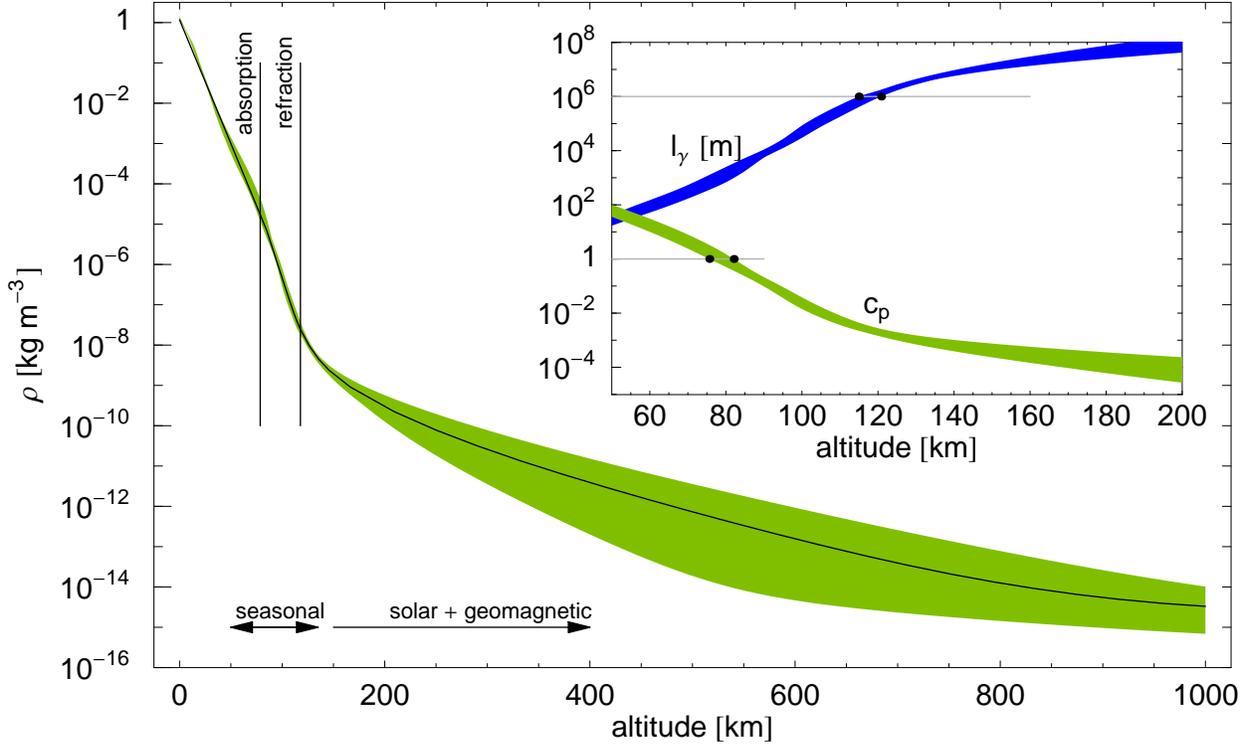}

\caption{\label{fig:atm} Total mass density $\rho$ as function of
  altitude. The black line shows the default density profile used
  throughout this paper and corresponds to the medium solar activity
  case recommended in~\cite{ecss}. The green/gray band denotes the
  maximal excursions from this default density profile predicted by
  the NRLMSISE-00 atmospheric model~\cite{nrlmsise}. The inset shows
  the photon mass induced minimal oscillation length $l_\gamma$
  (blue/dark band) and the absorption coefficient $c_p$ (green/light
  band) as functions of the altitude in the relevant range. The bands
  are due to full variation of density profiles shown in the main
  plot. The arrows at the bottom denote the main cause for the density
  variation in that altitude range.}
\end{figure}

The Earth atmosphere is roughly compromised of three layers: the
homosphere from $0-90\,\mathrm{km}$, the thermosphere starting at
$90\,\mathrm{km}$ and ranging up to $250-400\,\mathrm{km}$ depending on
solar and geomagnetic activity, as well as the exosphere which begins at the
top of thermosphere and extends into space. In the homosphere, winds
and turbulence mix all species very well and thus the composition is
independent of altitude. In the thermosphere turbulent mixing ceases
to be effective and the different species start to diffuse separately.
This diffusion is driven by gravitation and thermal gradients. In the
exosphere, finally, the mean free path of the lighter atoms like
hydrogen becomes large enough such that they can escape to space.  In
the thermosphere solar energy is absorbed by the photo-dissociation of
molecular oxygen. Thus, a sizable amount of free, atomic oxygen
appears and our assumption of constant composition fails. However, in
terms of electron density, 2 oxygen atoms have the same number of
electrons as 1 oxygen molecule, therefore the total mass density
is still a very precise indicator of the electron density\footnote{We
  verified that this holds to better than 1\% in the relevant altitude
  range.}.

There are no weather phenomena, in the ordinary sense, in the
thermosphere, nonetheless its density does depend on various variable
factors. This density distribution depends on the energy input from the
Sun via extreme ultraviolet light (EUV) and due to direct heating by
charged particles from cosmic radiation and solar wind. As a result,
the density depends on the amount of solar EUV radiation, which itself
depends on geographic latitude, the time of the year and the apparent
local solar time. Also the solar activity and geomagnetic activity as
well as the location within the geomagnetic field do have a
non-negligible influence. For a general overview
see~\cite{Jacchia,ecss}. Direct measurements of the total mass density
in the altitude range from $150-200\,\mathrm{km}$ can be performed by
the observation of the decay rate of very low altitude satellite
orbits. These measurements indicate densities of a few times
$10^{-9}\,\mathrm{kg}\,\mathrm{m}^{-3}$ at $150\,\mathrm{km}$. The
observed diurnal variations are 25\%, whereas the observed seasonal
variation is somewhat larger with 40\%~\cite{king,bowman}.

In~\cite{ecss}, the MSISE-90~\cite{msise} model\footnote{MSISE stands
  for 'mass spectrometer and incoherent scatter, extended'.} is recommended for
use in space missions, with three average density profiles from
$0-900\,\mathrm{km}$ corresponding to three different levels of solar
and geomagnetic activity. We use the one called medium activity,
corresponding to $F_{10.7}=F_{10.7}^\mathrm{avg}=140$ and $A_p=15$.
Where, $F_{10.7}$ denotes the $\lambda=10.7\,\mathrm{cm}$ flux density
from the Sun in units of $10^{-22}\,\mathrm{J}\,\mathrm{m}^{-2}$.
$F_{10.7}^\mathrm{avg}$ is the 81 day average of $F_{10.7}$.  The
$F_{10.7}$ index closely traces the solar UV emissions and the sun
spot number. $A_p$ is the daily average of the $a_p$ index, which
measures the perturbation of the geomagnetic field in units of
$2\,\mathrm{nT}$, see section~\ref{sec:geomagnet}. The corresponding
density profile is shown as black line in figure~\ref{fig:atm}.

In order to study the impact of variations of the density profile we
use an updated version of MSISE-90, the so called
NRLMSISE-00~\cite{nrlmsise} model. The differences between the two
models for the same set of input parameters, are however small.
NRLMSISE-00 takes as input the day of year, the local apparent solar
time, the geodetic latitude and longitude, $F_{10.7}$ and
$F_{10.7}^\mathrm{avg}$ and $A_p$ (or a series of average values of
$A_p$). In order to to estimate the maximal possible density
excursions, we varied: $F_{10.7}$ and $F_{10.7}^\mathrm{avg}$ jointly
from 40 to 380\footnote{This covers the extremes during one full
  11-year solar cycle, according to~\cite{ecss}.}, $A_p$ from 0 to
100\footnote{In principle, $A_p$ can reach values of up to 300 during
  the strongest geomagnetic storms.  These times, would however have
  to be discarded anyway since the fidelity of geomagnetic model can
  not be ensured at these times.  Note, that the density variations
  caused by $A_p>100$ especially affect the vicinity of
  $120\,\mathrm{km}$, {\it i.e.}  precisely the regions where
  $\eta_\gamma$ will be located.}, longitude from $0^\circ$ to
$90^\circ$\footnote{The $-90^\circ$ to $0^\circ$ range just swaps
  result between summer and winter.}, the day of the year from 1 to
365, the local apparent solar time from
$0\,\mathrm{h}-24\,\mathrm{h}$. For each altitude we determined the
minimal and maximal value of density due to all these different input
parameters, the result is shown as the green/gray band in
figure~\ref{fig:atm}. In the altitude range from $50-135\,\mathrm{km}$
seasonal changes and the geodetic latitude have the greatest effect,
wheres for higher altitude the main effects are due solar and
geomagnetic activity.  These regions are indicated by black arrows in
figure~\ref{fig:atm}.  The obtained values for density variation agree
well with the ones found in~\cite{king,bowman}. The inset in
figure~\ref{fig:atm} shows how $\eta_\mathrm{esc}$ and $\eta_\gamma$
change due to those density variations. The values and ranges are:
\beq
\eta_\mathrm{esc}=78^{+4}_{-3}\,\mathrm{km}\quad\mathrm{and}\quad
\eta_\gamma=118 \pm 3\,\mathrm{km}\,.
\eeq
We see that the limiting factor is indeed refraction and not
absorption and both factors need to be included for an accurate
calculation. Assuming an axion conversion path of around
$1000\,\mathrm{km}$, this is less than a 1\% change. We actually
verified that the GECOSAX flux does not change by more than 5\% due
atmospheric density variations, therefore the inclusion of atmospheric
effects via an average density profile is fully warranted.

\subsection{Axion conversion probability}

The probability for axion-photon conversion including the full path
and medium dependence is given by~\cite{vanBibber:1988ge}
\beq
\label{eq:probability}
P_{a\gamma}(m_a,\omega,t)= {\cal A}_t
\left|\int_0^{\alpha_t} d\lambda^\prime\,
B_\perp^\alpha(\lambda^\prime, t)\cdot
\exp\left\{i \int_0^{\lambda^\prime} d\lambda^{\prime\prime}\,
\frac{1}{2}\left[\frac{m_\gamma\left[h_{P_t}(\lambda^{\prime\prime})
\right]^2 - m_a^2}{\omega} - i \Gamma\left[h_{P_t}(\lambda^{\prime\prime})
\right]\right]\right\}\right|^2\,,
\eeq
with
\beq
\label{abs} {\cal A}_t\equiv \frac{g_{a\gamma}^2}{4}\,
\exp\left\{-\int_0^{\alpha_t} d\lambda\,
  \Gamma\left[h_{P_t}(\lambda)\right]\right\}\, \eeq and \beq
B_\perp^\alpha(\lambda^\prime, t)\equiv
\left|\vec{B}\left[\vec{p}_t(\lambda^\prime)\right]\times
  \vec{e}_\alpha^{\,t}\right|.  \eeq Here, $\omega$ is the axion
energy. The time dependence of $P_{a\gamma}$ is entirely due to change
of the geometry with time as explained in section~\ref{sec:orbit}. For
each time $t$ we solve for the position of the satellite $\vec{X}_t$
and for the position of the Sun $\vec{S}_t$. This information is used
to derive $\vec{p}_t(\lambda)$, the parametric form of the axion path
or the line of sight as defined in Eq.~\ref{eq:los}. The quantity
$\lambda$ parametrizes the position along the line of sight and
$\alpha_t$ denotes the length of the line of sight for the time $t$;
we have $\lambda\in[0,\alpha_t]$. $\Gamma$ and $m_\gamma$ only depend
on the density of the atmosphere which itself is a functions of the
height above mean sea level $h_{P_t}(\lambda)$. Only $\vec{B}$ has a
complete dependence on the position vector $\vec{p}$. For the various
necessary coordinate transformations we refer the reader to
appendix~\ref{app:coords}.

The integral in Eq.~(\ref{eq:probability}) has no closed form solution
and therefore has to be integrated numerically. For the numerical
integration a problem arises at very low altitudes, where the air
density is high and hence $m_\gamma$ is large. This gives rise to
extremely fast oscillation of the integrand of the innermost integral,
{\it i.e.} the integration with respect to $\lambda^{\prime\prime}$
becomes practically impossible for sufficiently small heights.  On the
other hand, $\Gamma$ also becomes very large and thus those parts of
the path where $m_\gamma$ is very large do not contribute to the
transition amplitude.  The solution is to reverse the direction of
integration by substituting $\lambda$ with $\alpha_t-\lambda$ and at
the same time all integrals now run from $\alpha_t$ to $0$. Next, the
integral is partitioned using a simple bisection: First the integral
from $\alpha_t$ to $\alpha_t/2$ and then from $\alpha_t/2$ to
$\alpha_t/4$ \ldots, until the contribution of the last part evaluated
is smaller than a preset precision goal, in our case this is $\Delta
P/P=10^{-5}$.

\subsection{Solar axion flux}
\label{sec:flux}

The Sun produces axions throughout its whole interior, although the
hottest regions with the highest photon density contribute the bulk of
the axion production. Since the angular size of the Sun and the axion
producing region are non-negligible compared to the typical angular
resolution of x-ray telescopes, we cannot treat the Sun as point
source of axions. A detailed study of solar surface axion luminosity
has been presented in~\cite{Andriamonje:2007ew}. Its results have been
made accessible to us in machine readable format by one of the 
authors~\cite{georg}. We will denote the solar surface axion
luminosity by $\varphi_a(r,E)$, following the notation 
in~\cite{Andriamonje:2007ew}, where $r$ is the dimensionless fraction
of the solar radius. Thus, the solar axion flux spectrum produced by
the Sun up to a certain radius $r_s$ is obtained by
\begin{equation}
\left.\frac{d\Phi_\odot}{dE}\right|_{r_s}=2\pi\int_0^{r_s}\,dr\,r\,\varphi_a(r,E)\,.
\end{equation}
The usually quoted solar axion flux assumes $r_s=1$ and can be written
as~\cite{Andriamonje:2007ew}
\begin{equation}
\left.\frac{d\Phi_\odot}{dE}\right|_{r_s=1} = 6.02 \cdot 10^{10} E^{2.481} \cdot
e^{-E/1.205}  \left(\frac{g_{a\gamma}}{10^{-10}\,
\mathrm{GeV}^{-1}}\right)^2\,\mathrm{cm}^{-2}\,\mathrm{s}^{-1}\mathrm{keV}^{-1}\,.
\end{equation}
Throughout this work, unless otherwise stated, we use
$g_{a\gamma}=10^{-10}\,\mathrm{GeV}^{-1}$. Note that the vast majority
of this flux originates within the inner $20\%$ of the solar
radius, {\it i.e.} $r_s=0.2$. Since the background will be
proportional to $r_s^2$, the signal significance will not be optimal
for $r_s=1$, but for some smaller value. This issue is studied in
detail in appendix~\ref{app:signal} and we adopt $r_s=0.13$. The flux
we are using is then given by
\begin{equation}
\label{eq:solarflux}
\frac{d\Phi_\odot}{dE}\equiv\left.\frac{d\Phi_\odot}{dE}
\right|_{r_s=0.13}=1.72 \cdot 10^{10} E^{3.210} \cdot
e^{-E/1.135}  \left(\frac{g_{a\gamma}}{10^{-10}\,
\mathrm{GeV}^{-1}}\right)^2\,\mathrm{cm}^{-2}\,\mathrm{s}^{-1}\mathrm{keV}^{-1}\,.
\end{equation}
The resulting axion fluence in the energy range from
$1-10\,\mathrm{keV}$ is
$3.58\cdot10^{11}\,\mathrm{axions}\,\mathrm{cm}^{-2}\,\mathrm{s}^{-1}$
for $r_s=1$ and it is
$2.21\cdot10^{11}\,\mathrm{axions}\,\mathrm{cm}^{-2}\,\mathrm{s}^{-1}$
for $r_s=0.13$. Also, the mean axion energy changes from
$4.2\,\mathrm{keV}$ for $r_s=1$ to $4.8\,\mathrm{keV}$ for
$r_s=0.13$. In appendix~\ref{app:signal}, it is demonstrated that the
loss of about $40\%$ in signal is compensated by a large decrease in
background.

From a comparison of the results obtained in~\cite{GGR} and
in~\cite{Andriamonje:2007ew}, it is estimated that this flux
calculation is accurate to about $5\%$, due to changes in the
underlying solar model.  In our analysis, we will also neglect the
annual variation of the Sun-Earth distance which causes a $5\%$
modulation of the signal.

\section{Resulting x-ray fluxes}
\label{sec:fluxes}

\subsection{General orbits}

\begin{figure}[t!]
\includegraphics[width=0.7\textwidth]{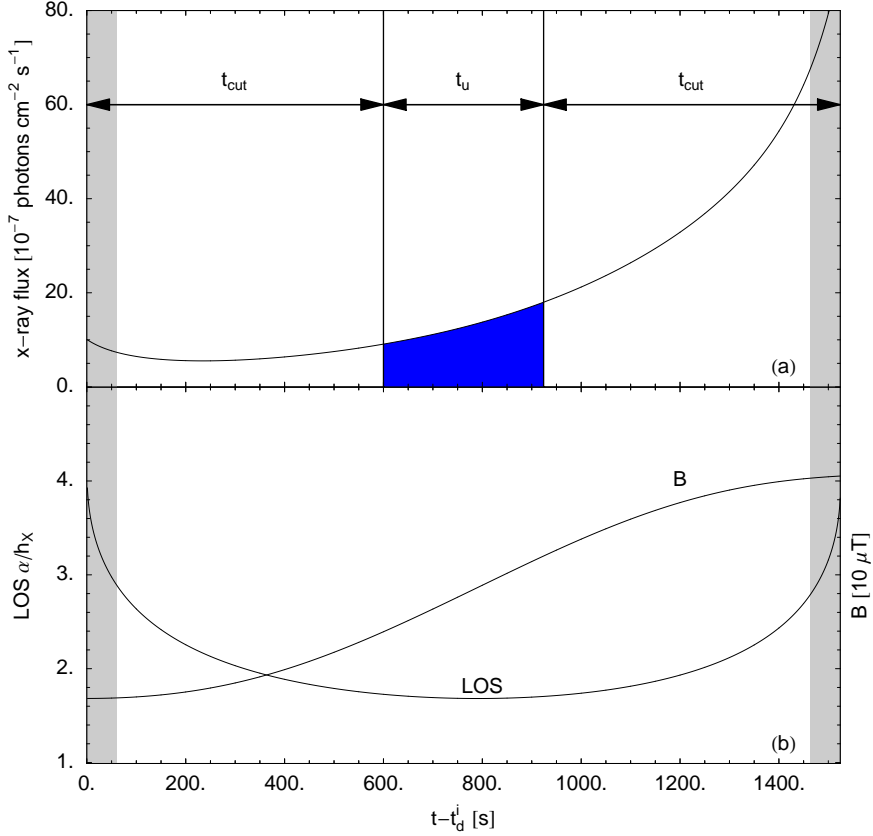}
\caption{\label{fig:typorb} A typical orbit for satellite 25399.
  $t_d^i$ is 2007-12-31, 23:48:33 UTC and the duration of the dark
  orbit is $1524\,\mathrm{s}$. Panel (a) shows the resulting x-ray
  flux from GECOSAX integrated over the energy range from
  $1-10\,\mathrm{keV}$. The blue shaded area gives the useful
  integrated GECOSAX flux after accounting for $t_\mathrm{cut}$.
  Panel (b) shows the length of the axion conversion path in units of
  the actual altitude of the satellite (line labeled LOS) and the
  total magnetic field strength at the location of the satellite B in
  units of $10\,\mu\mathrm{T}$ (line labeled B). The gray shaded areas
  are the first and last $60\,\mathrm{s}$ of the dark orbit.}
\end{figure}

We now have all the tools at hand to study the GECOSAX effect in detail
for any given satellite. Before we delve into finding the optimal
orbits in the following section, we briefly describe the GECOSAX flux
along a typical orbit. This will help to clarify some notation and to
give a basic overview of the issues involved. Figure~\ref{fig:typorb}
shows one dark orbit, {\it i.e.} that part of the orbit which is in
the Earth shadow, of satellite 25399 beginning at 2007-12-31, 23:48:33
UTC and lasting $1524\,\mathrm{s}$. We will call the beginning of a
dark orbit $t_d^i$ and the end $t_d^f$. The duration of the dark orbit
is then given by $t_d^f-t_d^i$. Since the satellite may not be able to
start observation immediately at $t_d^i$, we may have to cut away some
time at the beginning and end of the dark orbit, called
$t_\mathrm{cut}$; thus the useful duration of the orbit is
$t_u=t_d^f-t_d^i-2t_\mathrm{cut}$. Generally, the duration of dark
orbits will vary throughout the year and is different from orbit to
orbit. For some parts of the year, there may even be no dark orbits at
all, quite analogous to the fact that during summer the Sun does not
set north of the polar circle. The integrated useful GECOSAX flux for
each dark orbit $o$, which corresponds to the blue shaded area in
panel (a) of Fig.~\ref{fig:typorb}, is given by

\begin{equation}
\label{eq:gcxflux}
\langle \Phi\rangle_\mathrm{gcx}=
\int_{t_d^i+t_\mathrm{cut}}^{t_d^f-t_\mathrm{cut}}\,dt'
\int_{1\,\mathrm{keV}}^{10\,\mathrm{keV}}\,dE'\,
  P_{a\gamma}(t',E')\frac{d\Phi_{\odot}}{dE}(E')\,,
\end{equation}
where $t_d^{i/f}$ and $P_{a\gamma}(t',E')$ are different from orbit to
orbit and have to be computed correspondingly.

Panel (b) of Fig.~\ref{fig:typorb} shows that the axion conversion
path is always considerably longer than the altitude of the satellite,
its length ranges from about 1.6 times up to 4 times the altitude of
the satellite. This fact is responsible for most of the signal
increase with respect to our previous estimate
in~\cite{Davoudiasl:2005nh}. Also, within the first and last few
$10\,\mathrm{s}$ of the dark orbit there is large variation in this
length: nearly one half of the total path length variation happens
within the first and last $60\,\mathrm{s}$ of the dark orbit.
Therefore, in order to be not overly sensitive to errors in timing and
positioning we exclude those first and last $60\,\mathrm{s}$ from the
analysis. 
This is indicated by the gray shaded areas in
Fig.~\ref{fig:typorb}. We also see that the variation of the
magnetic field is non-negligible. This will depend strongly on the
path of the satellite with respect to the geomagnetic field. The orbit
of satellite 25399 has an inclination of $98^\circ$ and thus the
satellite does traverse the region of the geomagnetic poles; in this
case it is the south geomagnetic pole. Also, Fig.~\ref{fig:typorb}
shows that the orientation of the magnetic field is nearly parallel to
the axion path at the beginning of the dark orbit since the very large
path length right at beginning does not cause a corresponding increase
in GECOSAX flux. At the end of the dark orbit the magnetic field has a
larger component perpendicular to the axion path and hence the
increase in path length is well reflected in a corresponding increase
in the GECOSAX flux.

\subsection{Optimal orbits}

In this section we will apply the formalism developed in the previous
sections to determine what constitutes an optimal orbit for observing
GECOSAX.  In some approximation, the signal is proportional to
$B^2L^2$, therefore we would like to have orbits which have the
maximum possible path length in the highest possible magnetic field.
This points to high altitude satellites traversing the region of
geomagnetic poles.  This requires inclinations greater than about
$70^\circ$. Instead of designing an optimal orbit, which given the
many free parameters is a daunting task, we took a sample of existing
orbits, {\it i.e.} orbits which actually are used or have been used
for real scientific satellite missions. We took the TLEs of 50
satellites with apogees below $1000\,\mathrm{km}$ from~\cite{tle}.
Nearly, all of these orbits have low eccentricity $<0.1$. The apogees
of these satellites have an approximately Gau\ss ian distribution with
a mean of $650\,\mathrm{km}$ and a standard deviation of about
$120\,\mathrm{km}$. The inclinations are strongly clustered around
$80^\circ$; 28 satellites have inclinations in the range
$70^\circ-90^\circ$. Thus, this sample seems to be well suited for our
study. The US SPACECOM identification number and the number of the TLE
set used are given in Table~\ref{tab:tle}. Our goal here is solely to
determine the most suitable orbit and not the most suitable mission,
which is the combination of satellite and orbit. Thus, in the
following if we speak, say, of satellite 25544\footnote{This is the
  International Space Station (ISS).} we actually just refer to its
orbit and not the instruments or the satellite itself.

In determining the optimal orbit we need to distinguish two different
observational strategies, called `turning mode' and `fixed mode'.  A
turning mode satellite needs to avoid direct exposure of its x-ray
detection system to the sunlight (visible and x-ray) in order to
prevent any permanent damage. Typically, this Sun avoidance angle is
about $30^\circ$ and maximum sustained slew rates of
$6^\circ\,\mathrm{min}^{-1}$ have been demonstrated. Thus such a
satellite enters the Earth shadow pointing $30^\circ$ away from the
Sun, then it needs $30/6=5\,\mathrm{min}$ to turn into observation
position pointing to the Sun. Since these numbers may vary from
mission to mission and some safety margin will be necessary, we will
discard the first 10 minutes at entry into the Earth shadow as well as
the last 10 minutes prior to exit of the Earth shadow, thus giving up
20 minutes of each orbit. Fixed mode satellites have instrumentation
which can withstand direct irradiation by the Sun, or protective
shields that can be deployed quickly (compared to slew time), and thus
can do their maneuvering to point to the Sun in the bright parts of
their orbit.  Therefore, in principle, fixed mode satellites can use the entire
part of their orbit in the Earth shadow. We have noted previously that
the precise time of entry or exit of the Earth shadow are somewhat
uncertain due to geometrical and refractive effects. Also the axion
conversion rate has its peak values at the entry and exit points where
its time derivative is largest. Combining these factors, the GECOSAX
rate within first and last few $10\,\mathrm{s}$ of each dark orbit
have fairly high uncertainties. Therefore, we will exclude the first
and last 60 seconds of each dark orbit from the analysis even for
fixed mode satellites.  This is by no means a technical necessity but
is a conservative choice to ensure reliability of our results.

For each of the satellites in Table~\ref{tab:tle} we computed the
integrated, energy averaged GECOSAX flux $\langle
\Phi\rangle_\mathrm{gcx}$ for each orbit (about 5500 per satellite)
from January 1st 2008 till December 31st 2008 with time steps of
$60\,$s\footnote{$t_i^d$ and $t_f^d$ were determined to within
  $1\,\mathrm{s}$; `$60\,\mathrm{s}$' refers to the integration time
  step for computing the average flux.}. $\langle
\Phi\rangle_\mathrm{gcx}$ is computed in the limit of
$m_a\rightarrow0$. $\langle \Phi\rangle_\mathrm{gcx}$ accounts for the
time cut away at both ends of the dark orbit; as a result $\langle
\Phi\rangle_\mathrm{gcx}$ is different for turning and fixed
observation modes. The signal per unit area for a given orbit $o$ is
then given by $\sigma_o=\langle\Phi\rangle_\mathrm{gcx}^o \times
t_u^o$.  In reality there will be some background $b$ as well to the
measurement and thus the pertinent quantity to optimize is not the
total signal $s$ but signal divided by the square root of the
background $s/\sqrt{b}$.  We call this quantity significance. The
signal is given by $s_o=\sigma_o \times A$, where $A$ is the area of
the detector. The background is given by $b_o=F\times A \times t_u^o$,
where $F$ is the background rate per unit area. $F$ here is to be
understood as the background rate $f$ integrated over x-ray energies
from $E_\mathrm{min}=1\,\mathrm{keV}$ to
$E_\mathrm{max}=10\,\mathrm{keV}$ and over the solid angle covered by
the axion producing region in the Sun. Assuming a uniform distribution
of the background in energy and solid angle, we obtain the following
relation $F=\Omega_s (E_\mathrm{max}-E_\mathrm{min}) f$. As mentioned
in section~\ref{sec:flux} and explained in detail in
appendix~\ref{app:signal}, we take the signal producing region to be
$0.13R_\odot$. The Sun's angular diameter is $32'$, thus the solid
angle $\Omega_s$ subtended by the signal producing region is
$\Omega_s=\pi(0.13\cdot32/2)^2\,\mathrm{arcmin}^2=13.6\,\mathrm{arcmin}^2$.
Thus we obtain
\begin{equation}
\label{eq:frate}
F=122.4 \left(\frac{f}{\mathrm{s}^{-1}\,\mathrm{cm}^{-2}\,
\mathrm{keV}^{-1}\,\mathrm{arcmin}^{-2}}\right)\,
\mathrm{s}^{-1}\,\mathrm{cm}^{-2}\,.
\end{equation}

Next, we define the unit significance for a single orbit $o$ like this
\begin{equation}
\label{eq:sigQ}
S^o\equiv\frac{t_u^o A \langle\Phi\rangle_\mathrm{gcx}^o}
{\sqrt{t_u^o A F}}=\underbrace{A^{1/2}F^{-1/2}}_{\equiv Q}\,
\underbrace{t_u^{1/2}\langle\Phi\rangle_\mathrm{gcx}^o}_{\equiv\Sigma_o}=Q\Sigma_o\,.
\end{equation}
$Q$ is called quality factor and does not depend on a particular orbit
but only on the instrument used for x-ray observation\footnote{The 
background rate $F$ actually has some dependence on the position 
relative to the Earth and is not constant in time. We will comment 
on this point later in more detail.}, whereas $\Sigma_o$ is determined by
the orbit itself and the observation mode. Using
Eq.~(\ref{eq:frate}), we can express $Q$ in terms of $f$, which
yields
\begin{equation}
Q=0.09\,\mathrm{cm}^2\,\mathrm{s}^{-1/2}\,\left(\frac{A}{\mathrm{cm}^2}\right)^{1/2}\left(\frac{\mathrm{s}^{-1}\,\mathrm{cm}^{-2}\,\mathrm{keV}^{-1}\,\mathrm{arcmin}^{-2}}{f}\right)^{1/2}\,.
\end{equation}

We now sort all orbits of
a particular satellite in decreasing order of $\Sigma_o$ and add the
first $I$ orbits from the top of the list to the analysis. Thus we can
compute the total significance $S$ for $I$ orbits:
\begin{equation}
\label{eq:sigma}
S(I)\equiv\left(\sum_{i=1}^{I} (S^o)^2\right)^{1/2}
=\left(\sum_{i=1}^{I} (\Sigma_o Q)^2\right)^{1/2}
=Q\underbrace{\left(\sum_{i=1}^{I} \Sigma_o^2\right)^{1/2}}_{\equiv\Sigma}
=Q\Sigma
\quad\mathrm{with}\quad
t_u=\sum_{i=1}^{I} t_u^i\,.
\end{equation}
The maximal possible value for $I$ is given by the number of orbits in
the covered time period, which in our case is one year. The definition
of $S$ is inspired by the form of a Gau{\ss}ian $\chi^2$-function and
the fact that for 1 degree of freedom, a $\chi^2$ difference of $x$
corresponds to $\sqrt{x}\,\sigma$ significance. In this way, the
contribution from a particular instrument, encoded in the quality
factor $Q$, and of a particular orbit, encoded in $\Sigma$, can be
cleanly separated. The only remaining effect of the particular
satellite on $\Sigma$ is the necessary $t_\mathrm{cut}$, where fixed
and turning mode should bracket most realistic setups.

\begin{figure}[t!]
\includegraphics[width=0.7\textwidth]{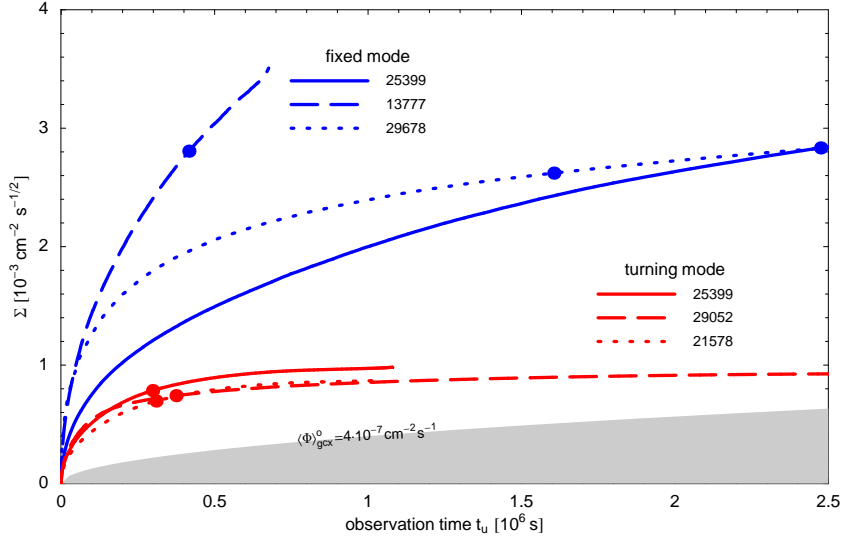}
\caption{\label{fig:sig} Shown is the significance $\Sigma$ as defined
  in Eq.~\ref{eq:sigma} as a function of the observation time $t_u$.
  In blue those three satellites are shown for which $\Sigma$ reaches
  the largest possible value in fixed mode. Whereas, in red the
  three satellites are shown for which $\Sigma$ reaches the largest
  possible value in turning mode. The big dots denote the times at
  which each satellite has reached 80\% of its maximal $\Sigma$ in the
  corresponding mode. The gray shaded area shows $\Sigma$ for the case
  where the average flux per orbit is constantly
  $\langle\Phi\rangle^o_\mathrm{gcx}=4\cdot10^{-7}\,\mathrm{cm}^{-2}\,\mathrm{s}^{-1}$,
  which corresponds to the result obtained
  in~\cite{Davoudiasl:2005nh}.}
\end{figure}

For each value of $I$ we obtain a value for
$\Sigma$ and $t_u$ and we can plot $\Sigma$ as a function of $t_u$.
This is shown in Fig.~\ref{fig:sig} for those 3 satellites which
have the largest maximal obtainable $\Sigma$ in turning (red lines)
or fixed mode (blue lines).
In case all orbits have a very similar value of $S_o$, $\Sigma$ is
approximately proportional to $\sqrt{t_u}$. The fixed mode satellite
25399 exhibits this type of behavior for all times shown in
Fig.~\ref{fig:sig}.  However, for some satellites (like 13777) all
available orbits are used up at relatively small values of $t_u$.
Obviously, satellites which do not have to turn, {\it i.e.} the fixed
mode ones, have a clear advantage. They can accumulate more useful
time since they do not lose time by turning once they enter the dark
orbit. Moreover, as indicated in Fig.~\ref{fig:typorb}, the GECOSAX
fluxes are highest either at the very end or beginning of each dark
orbit. Thus, they can reach values of $\Sigma=0.0035
\,\mathrm{cm}^{-2}\,\mathrm{s}^{-1/2}$, about 3 times larger than
turning mode satellites. This factor of 3 is very relevant as $S$ only
increases as the square root of time, area or the inverse background
rates, therefore a 3 times larger $\Sigma$ allows a 9 times smaller
area or 9 times larger background while having the same statistical
significance.

For turning mode satellites (red lines), there are a few long
duration, high flux orbits which contribute the bulk to $\Sigma$ and
afterward the curve increases much more slowly than $\sqrt{t_u}$. Thus for
an optimal use of resources, it is advisable to avoid those orbits and
to restrict the axion search to only the best available orbits.

Therefore, we introduce the quantity
$\Sigma_{80}\equiv0.8\Sigma_\mathrm{max}$, which is just 80\% of the
maximal obtainable $\Sigma$. The reduction in significance is small
compared to the savings in observation time when the experiment is
restricted to those orbits which allow to reach $\Sigma_{80}$.
$\Sigma_{80}$ is marked by a dot on each curve in
Fig.~\ref{fig:sig}. The corresponding times $t_{80}$ are much shorter
than the maximal available $t_u$; in case of satellite 29052 the
reduction is nearly a factor 10 with a minimal sacrifice in $\Sigma$.
We see that $t_{80}$ can be significantly below $10^6\,\mathrm{s}$ for
the best available satellites. For example satellite 13777 in fixed
mode would reach a
$\Sigma_{80}=0.0028\,\mathrm{cm}^{-2}\,\mathrm{s}^{-1/2}$ in about 1
week\footnote{Including the 2 times $60\,\mathrm{s}$ per orbit.}. The
values of $\Sigma_{80}$, the total time needed and the number of
necessary orbits in both fixed and turning mode are listed for all
considered satellites in Table~\ref{tab:tle}.

All optimal orbits we identified fulfill the naive expectation stated
at the beginning of this section: they maximize $B^2L^2$ by having
their dark orbits in the regions of the strongest geomagnetic field
and they all have orbits close to the maximum altitude of
$1000\,\mathrm{km}$ considered here. For the sake of completeness, we
have added the performance of the current ISS orbit in
Table~\ref{tab:tle}. Due to the very low altitude of the ISS of around
$380\,\mathrm{km}$, this orbit does not perform very well. On the
other hand, if the restriction on the maximal altitude is relaxed,
completely new types of orbits become available.

One interesting such class
are the so called Molniya orbits, which are highly eccentric with
perigees $\sim1000\,\mathrm{km}$ and apogees
$\sim40\,000\,\mathrm{km}$. These orbits have a period of
$12\,\mathrm{h}$ and have a repeat ground track, {\it i.e.} they reach
the same point above the Earth every $12\,\mathrm{h}$ and can thus
cross the geomagnetic pole every second orbit. Due to their primary
design goal of allowing communication with high latitudes in Russia, they
have their perigee on the Southern hemisphere, some of them very
close to geomagnetic South pole. This implies that during antarctic
night, which corresponds to the summer months on the northern
hemisphere, a satellite on such an orbit would be in darkness in a
very high $B$-field region at an altitude of $\sim1000\,\mathrm{km}$
every $24\,\mathrm{h}$ for a duration of about $1\,\mathrm{h}$.
Therefore, a few of the best available such orbits out of 24
tested ones are listed as well in Table~\ref{tab:tle}. Note that
these orbits could reach comparable sensitivities to the best
available ones discussed so far. Especially, in turning mode they could
yield a significance 20\% better than any other orbit. The total
observation time needed would be very short, on the order of few
$10^5\,\mathrm{s}$. However, the variability of the Earth magnetic
field so far out is greater and would require a more careful
consideration than the present note allows for. Also, the obtained
bound due to the much longer average length of the axion path, would
deteriorate at a smaller value of $m_a$ compared to the calculation
presented in figure~\ref{fig:sens}.


\section{Sensitivity to $\mathbf{g_{a\gamma}}$}
\label{sec:sens}

\begin{figure}[t!]
\includegraphics[width=0.5\textwidth]{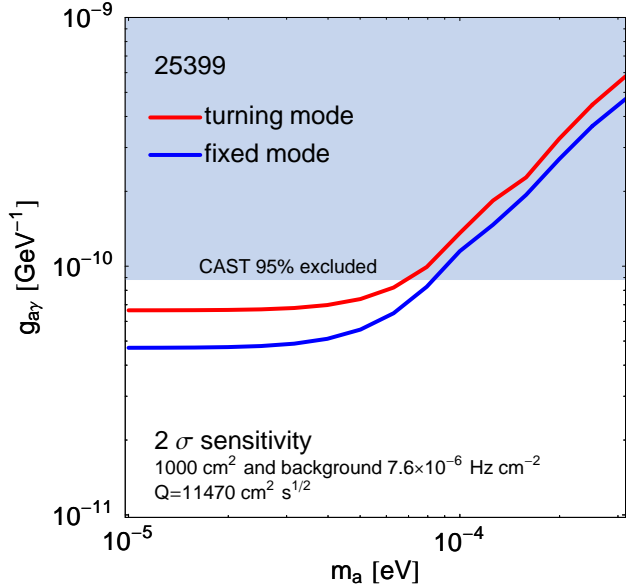}
\caption{\label{fig:sens} Sensitivity to $g_{a\gamma}$ as a function
  of the axion mass $m_a$ at $2\,\sigma$ (95\%) confidence level. The
  blue shaded region is excluded by the CAST
  experiment~\cite{Andriamonje:2007ew}.}
\end{figure}

In order to compute the sensitivity to $g_{a\gamma}$ we have to
specify a value or range of values for the quality factor $Q$.
Clearly, $Q$ is a very instrument-specific quantity and each existing x-ray
detector in space will have its unique value of $Q$. However, as shown
in Eq.~(\ref{eq:sigQ}), $Q=\sqrt{A/F}$ is a combination of two
factors: the effective x-ray collecting area and the background rate
$F$. In principle, these two factors can be scaled independently.
Therefore, we will consider the range of effective areas and range of
background rates $f$ found in real or planned x-ray satellite missions
separately. We list the effective area, the background rates
$(f, F)$ and the resulting quality factor $Q$ in
Table~\ref{tab:instruments}.
\begingroup
\squeezetable
\begin{longtable}{|ccc|rrrr|c|}
  \hline Mission&ID&Instrument&effective area&background rate
  $f$&background
  rate $F$&$Q$&reference\\
  &&&$\mathrm{cm}^2$&$10^{-8}\mathrm{cm}^{-2}\,\mathrm{s}^{-1}\,\mathrm{keV}^{-1}\,\mathrm{arcmin}^{-2}$&$10^{-6}\mathrm{cm}^{-2}\,\mathrm{s}^{-1}$&$\mathrm{cm}^2\,\mathrm{s}^{1/2}$&\\
  \hline
  XMM\footnote{Not in low Earth orbit.}&25989&EPIC MOS&900&29&36&5000&\cite{suzaku,xmm}\\
  XTE&23757&PCA&7000\footnote{This number corresponds to the value at
    the beginning of the mission.}&-&3600&2260\footnote{The PCA is a
    non-imaging detector and hence there is no background rate $f$
    given. Since it covers the whole Sun, the axion flux for $r_s=1$
    has be to taken, which is 1.6 times larger than the one for
    $r_s=0.13$. This
    correction factor has been applied to the $Q$ value quoted here.}&\cite{suzaku,xte}\\
  SUZAKU&28773&XIS FI&250&6.3&7.6&5735&\cite{suzaku1}\footnote{The
    background cited is the measured value,
    while observing the dark side of the Earth.}\\
  XEUS\footnotemark[20]&-&&50000&120&147&18443&\cite{xeus}\\
  \hline

\caption{\label{tab:instruments} Values for effective area $A$, the
  background rate $f$ and the integrated background rate $F$ for
  various existing and planned x-ray observatories. For those missions
  already in space we list the US SPACECOM ID number. $F$ is
  integrated over the energy range $1-10\,\mathrm{keV}$ and the source
  size of $13.6\,\mathrm{arcmin}^2$. Given are also the resulting quality
  factor $Q$ and the reference for the information.}
\end{longtable}
\endgroup
Table~\ref{tab:instruments} is not intended to be an exhaustive
survey, but to indicate the possibilities of a few contemporary
missions.  The effective areas range over
$(250-50\,000)\,\mathrm{cm}^2$, the background rates $F$ span $(8 - 3\,600)
\times 10^{-6}\,\mathrm{cm}^{-2}\,\mathrm{s}^{-1}$, and the resulting
$Q$ values are in the range $(2\,200 -
18\,000)\,\mathrm{cm}^2\,\mathrm{s}^{1/2}$. Taking the extreme
combinations of the effective areas and background rates from this
table, the corresponding range of $Q$ is $(300-81\,100)
\,\mathrm{cm}^2\,\mathrm{s}^{1/2}$. For the sensitivity estimate
presented in Fig.~\ref{fig:sens}, an effective area of
$1\,000\,\mathrm{cm}^2$ and a background rate of $7.6\times
10^{-6}\,\mathrm{cm}^{-2}\,\mathrm{s}^{-1}$ is assumed, yielding
$Q=11\,471\,\mathrm{cm}^2\,\mathrm{s}^{1/2}$.  The value $7.6\times
10^{-6}\,\mathrm{cm}^{-2}\,\mathrm{s}^{-1}$ corresponds to the
background rate measured by SUZAKU while observing the dark side of
the Earth~\cite{suzaku1}, in the energy range $0.5-10\,\mathrm{keV}$.
Therefore, this constitutes a guaranteed upper bound on any
interfering x-ray luminosity from the dark side of the Earth, {\it
  i.e} down to this level the dark side of the Earth is certainly dark
in x-rays. Note that the instruments on board SUZAKU are among the
most sensitive ones for extended sources~\cite{suzaku}.

For the computation of the actual sensitivity to $g_{a\gamma}$ we will
replace the significance defined in Eq.~(\ref{eq:sigma}) by the
correct form of $\chi^2$-function. Since the count rates are very low
it is necessary to use the Poissonian form of the $\chi^2$-function,
see {\it e.g.}~\cite{pdg}. There is considerable variation in the
GECOSAX flux along a single dark orbit as is obvious from
Fig.~\ref{fig:typorb}. Therefore, the signal to noise ratio will
also vary greatly and hence this time dependence can be
exploited. Thus, the data to be fitted consist of the time series of
all time bins of $60\,\mathrm{s}$ which are in a dark orbit and are not
within $t_\mathrm{cut}$ of either $t_i^d$ or $t_f^d$ and belong to one
of the those orbits which comprise $\Sigma_{80}$.
\begin{eqnarray}
b_{o,i}(E_j)&=&F\,A\,\Delta t \Delta E\,,\nonumber\\
n_{\mathrm{the}}^{o,i}(E_j,m_a)&=&g_{10}^4\,\langle\Phi^o_i(E_j,m_a)\rangle_{\Delta
t,\Delta E}\, A
\,\Delta t +b_{o,i}(E_j)\,,\nonumber\\
n_\mathrm{obs}^{o,i}(E_j)&=&b_{o,i}(E_j)\,,
\end{eqnarray}
with $g_{10}$ being $g_{a\gamma}$ in units of
$10^{-10}\,\mathrm{GeV}^{-1}$. $\langle\ldots\rangle_{\Delta t,\Delta
  E}$ is the average over the energy interval $\Delta E$ and the time
interval $\Delta t$. Here, $\Delta E=1\,\mathrm{keV}$ and $\Delta
t=60\,\mathrm{s}$. Next, we define the $\chi^2$-function as follows
\begin{equation}
\label{eq:chi}
\chi^2(g_{10},m_a)=2\sum_{o=1}^{o_{80}}\sum_{i=1}^{i\Delta t<t_u^o}\sum_{j=1}^{9}
n_\mathrm{the}^{o,i}(E_j,m_a)-n_\mathrm{obs}^{o,i}(E_j) +
n_\mathrm{obs}^{o,i}(E_j) \ln \frac{n_\mathrm{obs}^{o,i}(E_j)}{n_\mathrm{the}^{o,i}(E_j,m_a)}
\end{equation}
The bound $g_b$ on $g_{10}$ or $g_{a\gamma}$ is found by requiring
that $\chi^2(g_b,m_a)=4$, thus the bound is at $2\,\sigma$ or 95\%
confidence level. This is repeated for many values of $m_a$. The
resulting sensitivities for the top performing satellites for both the
fixed and the turning mode are shown in Fig.~\ref{fig:sens} as a
function of $m_a$. It turns out that accidentally, the same satellite
performs best in both modes.
The asymptotic sensitivities for $m_a\rightarrow0$ are
$4.7\cdot10^{-11}\,\mathrm{GeV}^{-1}$ for fixed mode and
$6.6\cdot10^{-11}\,\mathrm{GeV}^{-1}$ for turning mode. For comparison
the CAST asymptotic bound is
$8.8\cdot10^{-11}\,\mathrm{GeV}^{-1}$~\cite{Andriamonje:2007ew}. Thus
a fixed mode observation would improve the CAST limit by about a
factor of 2, which is considerable given that the signal scales as
$g_{a\gamma}^4$, {\it i.e.} the actual performance in terms of flux
sensitivity is more than 12 times better than CAST. Given the fact
that the flux sensitivity scales only as square root of time, area and
the inverse background rate, the actual increase in time or area would
have to be 150-fold to reach this sensitivity.

One important issue for the validity of the result is how it
would change if we allow the background to have systematic errors and
a time variation. From a purely statistical point of view, we observe
that the number of background events per time and energy bin typically is
very small with a mean value of $\mathcal{O}(1)$ for the $F$ and $A$ chosen
here. Thus the purely statistical variation of the background in each
single bin is around $100\%$, {\it i.e.} any fully, between all time
bins, uncorrelated systematic variation of the background would have
to be of that order of magnitude to produce a visible effect. This
seems to be very unrealistic. On the other hand, a common mode change
of the background, {\it i.e.} an effective systematic variation of $F$
would be very difficult to accommodate since the data contains bins
with nearly no signal. Those bins severely constrain $F$. Therefore,
one would have to introduce a systematic error which closely mimics
the time dependence of the axion signal.

Note, however, that almost
all time dependence of the background is due to either variations in
the geomagnetic field, either by position or due {\it e.g.} solar wind
or due to changes of position with respect to the Sun. Thus, typically,
backgrounds will be high if the magnetic field is low.  The signal,
however, will be large when the magnetic field is high. This is a very
strong anti-correlation. Also, the background mostly depends on the
total field strength and not on the component perpendicular to the
axion path. The position of the satellite with respect to the Sun is
not very different for consecutive orbits and hence background
events do not mimic the predicted time dependence
of the signal. The strongest background
rejection, however, is due to the known direction of the signal.
Therefore, the same strategy as used by CAST can be applied here as
well. The image of the axion-producing core of the Sun in the focal
plane of the x-ray telescope will cover only a few pixels out of the
whole sensor. All the other pixels can be used to measure the
background {\it in situ}. The main systematics in that case would be
due to pointing errors and the width and shape of the point spread
function of x-ray optics. In CAST these errors are estimated to be
negligible~\cite{Andriamonje:2007ew}.

In comparing the result from a full time and energy binned $\chi^2$
analysis as defined in Eq.~\ref{eq:chi} and the one of the
simplified treatment using the significance $\Sigma$ as defined in
Eq.~\ref{eq:sigma}, we find that using $\Sigma$ underestimates the
flux sensitivity by about 50\%. The resulting error in $g_{a\gamma}$
is about 5\%. Thus using $\Sigma Q$ to estimate the obtainable
sensitivity for $m_a\rightarrow0$ seems to be a conservative
approximation with very reasonable accuracy. Therefore the value of
$\Sigma_{80}$ in Table~\ref{tab:tle} can be used for a prediction of
the potential of a given satellite or instrument. Using this procedure
the sensitivity at $N\,\sigma$ can be obtained by
\begin{equation}
\label{eq:limit}
\left(\frac{g_b}{10^{-10}\mathrm{GeV}^{-1}}\right)=\left(\frac{\Sigma Q}{N}\right)^{-1/4}\,.
\end{equation}
Taking the largest $Q=81\,100\,\mathrm{cm}^2\,\mathrm{s}^{1/2}$
encountered in the discussion of Table~\ref{tab:instruments} and the
largest value of
$\Sigma=3.5\times 10^{-3}\,\mathrm{cm}^{-2}\,\mathrm{s}^{-1/2}$ found
from Fig.~\ref{fig:sig} we get a hypothetical limiting sensitivity of
\beq
g_b=2.9\times 10^{-11}\,\mathrm{GeV}^{-1}.
\label{eq:hypsen}
\eeq


\section{Discussion and Outlook}
\label{sec:conc}

The overall accuracy of our GECOSAX flux prediction depends on the
accuracy of the geometric description, the magnetic field model, the
air density profiles and their effects on x-ray propagation, as well
as the numerical implementation of Eq.~(\ref{eq:probability}). In
section~\ref{sec:orbit}, we concluded that geometry related effects
are accounted for within about $\pm25\,\mathrm{km}$ or $\pm
5\,\mathrm{s}$. This in turn introduces less than $10\%$ error in the
GECOSAX signal prediction. The main sources for these errors are, our
purely geometric definition of the Earth shadow and our treatment of
TLEs and SGP4. The latter source would be absent in an actual
measurement. Also, the time of entry into the Earth shadow is actually
easily accessible via the telemetry data of the satellite, {\it e.g.}
electricity production in the solar panels should be a precise
indicator. Concerning the magnetic field modeling, we found in
section~\ref{sec:geomagnet} that B-field errors should be less than
$5\%$ resulting in at most a $10\%$ error on the signal. This
component may be difficult to improve even in a real experiment. In
section~\ref{sec:propagation}, we found that a static average
atmospheric model can be safely used without introducing more than
$5\%$ error on the GECOSAX flux. Numerical integrations and the
coordinate transformation should not contribute to the total error
budget. We verified our code against available analytical results
~\cite{vanBibber:1988ge}. We also note that there is an approximately
$5\%$ annual modulation of the solar axion flux due to the variation
in the Sun-Earth distance.  This effect was not accounted for in our
computations and modulates our computed signal at the same level.
However, this is not a source of uncertainty and is in fact a
predicted feature of the signal.  Thus, we find that the results for
the GECOSAX flux presented here should have an error not exceeding
about $15\%$, which in a real experiment may be reduced down to about
$10\%$.  Given the above considerations, low Earth orbit measurements
of GECOSAX provide a novel experimental avenue for going beyond the
current laboratory bounds on the axion-photon coupling, for axion
masses below $10^{-4}\,\mathrm{eV}$.  We hope that the analysis
presented here will help motivate future experimental efforts in this
direction.

\begin{acknowledgements}
  We would like to thank M.~Kuster for collaboration during the early
  stages of this work as well as comments on a draft version, and
  D.~McCammon for various useful discussions and the access he
  provided to the preliminary SUZAKU results. We would also like to
  thank G.~Raffelt for sharing numerical results on solar axion surface
  luminosity, and S.~Maus from NOAA for his clarifications on the WMM
  2005.  The work of H.D. is supported by the United States Department of Energy
 under Grant Contract DE-AC02-98CH10886.
\end{acknowledgements}


\bibliographystyle{apsrev}
\bibliography{./references}

\begin{thebibliography}{44}
\expandafter\ifx\csname natexlab\endcsname\relax\def\natexlab#1{#1}\fi
\expandafter\ifx\csname bibnamefont\endcsname\relax
  \def\bibnamefont#1{#1}\fi
\expandafter\ifx\csname bibfnamefont\endcsname\relax
  \def\bibfnamefont#1{#1}\fi
\expandafter\ifx\csname citenamefont\endcsname\relax
  \def\citenamefont#1{#1}\fi
\expandafter\ifx\csname url\endcsname\relax
  \def\url#1{\texttt{#1}}\fi
\expandafter\ifx\csname urlprefix\endcsname\relax\def\urlprefix{URL }\fi
\providecommand{\bibinfo}[2]{#2}
\providecommand{\eprint}[2][]{\url{#2}}

\bibitem[{\citenamefont{Peccei and Quinn}(1977{\natexlab{a}})}]{Peccei:1977hh}
\bibinfo{author}{\bibfnamefont{R.~D.} \bibnamefont{Peccei}} \bibnamefont{and}
  \bibinfo{author}{\bibfnamefont{H.~R.} \bibnamefont{Quinn}},
  \bibinfo{journal}{Phys. Rev. Lett.} \textbf{\bibinfo{volume}{38}},
  \bibinfo{pages}{1440} (\bibinfo{year}{1977}{\natexlab{a}}).

\bibitem[{\citenamefont{Peccei and Quinn}(1977{\natexlab{b}})}]{Peccei:1977ur}
\bibinfo{author}{\bibfnamefont{R.~D.} \bibnamefont{Peccei}} \bibnamefont{and}
  \bibinfo{author}{\bibfnamefont{H.~R.} \bibnamefont{Quinn}},
  \bibinfo{journal}{Phys. Rev.} \textbf{\bibinfo{volume}{D16}},
  \bibinfo{pages}{1791} (\bibinfo{year}{1977}{\natexlab{b}}).

\bibitem[{\citenamefont{Weinberg}(1978)}]{Weinberg:1977ma}
\bibinfo{author}{\bibfnamefont{S.}~\bibnamefont{Weinberg}},
  \bibinfo{journal}{Phys. Rev. Lett.} \textbf{\bibinfo{volume}{40}},
  \bibinfo{pages}{223} (\bibinfo{year}{1978}).

\bibitem[{\citenamefont{Wilczek}(1978)}]{Wilczek:1977pj}
\bibinfo{author}{\bibfnamefont{F.}~\bibnamefont{Wilczek}},
  \bibinfo{journal}{Phys. Rev. Lett.} \textbf{\bibinfo{volume}{40}},
  \bibinfo{pages}{279} (\bibinfo{year}{1978}).

\bibitem[{\citenamefont{Preskill et~al.}(1983)\citenamefont{Preskill, Wise, and
  Wilczek}}]{Preskill:1982cy}
\bibinfo{author}{\bibfnamefont{J.}~\bibnamefont{Preskill}},
  \bibinfo{author}{\bibfnamefont{M.~B.} \bibnamefont{Wise}}, \bibnamefont{and}
  \bibinfo{author}{\bibfnamefont{F.}~\bibnamefont{Wilczek}},
  \bibinfo{journal}{Phys. Lett.} \textbf{\bibinfo{volume}{B120}},
  \bibinfo{pages}{127} (\bibinfo{year}{1983}).

\bibitem[{\citenamefont{Abbott and Sikivie}(1983)}]{Abbott:1982af}
\bibinfo{author}{\bibfnamefont{L.~F.} \bibnamefont{Abbott}} \bibnamefont{and}
  \bibinfo{author}{\bibfnamefont{P.}~\bibnamefont{Sikivie}},
  \bibinfo{journal}{Phys. Lett.} \textbf{\bibinfo{volume}{B120}},
  \bibinfo{pages}{133} (\bibinfo{year}{1983}).

\bibitem[{\citenamefont{Dine and Fischler}(1983)}]{Dine:1982ah}
\bibinfo{author}{\bibfnamefont{M.}~\bibnamefont{Dine}} \bibnamefont{and}
  \bibinfo{author}{\bibfnamefont{W.}~\bibnamefont{Fischler}},
  \bibinfo{journal}{Phys. Lett.} \textbf{\bibinfo{volume}{B120}},
  \bibinfo{pages}{137} (\bibinfo{year}{1983}).

\bibitem[{\citenamefont{Turner}(1986)}]{Turner:1985si}
\bibinfo{author}{\bibfnamefont{M.~S.} \bibnamefont{Turner}},
  \bibinfo{journal}{Phys. Rev.} \textbf{\bibinfo{volume}{D33}},
  \bibinfo{pages}{889} (\bibinfo{year}{1986}).

\bibitem[{\citenamefont{Csaki et~al.}(2002)\citenamefont{Csaki, Kaloper, and
  Terning}}]{Csaki:2001yk}
\bibinfo{author}{\bibfnamefont{C.}~\bibnamefont{Csaki}},
  \bibinfo{author}{\bibfnamefont{N.}~\bibnamefont{Kaloper}}, \bibnamefont{and}
  \bibinfo{author}{\bibfnamefont{J.}~\bibnamefont{Terning}},
  \bibinfo{journal}{Phys. Rev. Lett.} \textbf{\bibinfo{volume}{88}},
  \bibinfo{pages}{161302} (\bibinfo{year}{2002}), \eprint{hep-ph/0111311}.

\bibitem[{\citenamefont{Raffelt}(1996)}]{GGR}
\bibinfo{author}{\bibfnamefont{G.~G.} \bibnamefont{Raffelt}},
  \emph{\bibinfo{title}{Stars as Laboratories of Fundamental Physics}}
  (\bibinfo{publisher}{The University of Chicago Press}, \bibinfo{year}{1996}),
  \bibinfo{edition}{2nd} ed.

\bibitem[{\citenamefont{Pirmakoff}(1951)}]{Pirmakoff:1951pj}
\bibinfo{author}{\bibfnamefont{H.}~\bibnamefont{Pirmakoff}},
  \bibinfo{journal}{Phys. Rev.} \textbf{\bibinfo{volume}{81}},
  \bibinfo{pages}{899} (\bibinfo{year}{1951}).

\bibitem[{\citenamefont{Sikivie}(1983)}]{Sikivie:1983ip}
\bibinfo{author}{\bibfnamefont{P.}~\bibnamefont{Sikivie}},
  \bibinfo{journal}{Phys. Rev. Lett.} \textbf{\bibinfo{volume}{51}},
  \bibinfo{pages}{1415} (\bibinfo{year}{1983}).

\bibitem[{\citenamefont{Andriamonje et~al.}(2007)}]{Andriamonje:2007ew}
\bibinfo{author}{\bibfnamefont{S.}~\bibnamefont{Andriamonje}}
  \bibnamefont{et~al.} (\bibinfo{collaboration}{CAST}), \bibinfo{journal}{JCAP}
  \textbf{\bibinfo{volume}{0704}}, \bibinfo{pages}{010} (\bibinfo{year}{2007}),
  \eprint{hep-ex/0702006}.

\bibitem[{\citenamefont{Yao et~al.}(2006)}]{Yao:2006px}
\bibinfo{author}{\bibfnamefont{W.~M.} \bibnamefont{Yao}} \bibnamefont{et~al.}
  (\bibinfo{collaboration}{Particle Data Group}), \bibinfo{journal}{J. Phys.}
  \textbf{\bibinfo{volume}{G33}}, \bibinfo{pages}{1} (\bibinfo{year}{2006}).

\bibitem[{\citenamefont{Davoudiasl and Huber}(2006)}]{Davoudiasl:2005nh}
\bibinfo{author}{\bibfnamefont{H.}~\bibnamefont{Davoudiasl}} \bibnamefont{and}
  \bibinfo{author}{\bibfnamefont{P.}~\bibnamefont{Huber}},
  \bibinfo{journal}{Phys. Rev. Lett.} \textbf{\bibinfo{volume}{97}},
  \bibinfo{pages}{141302} (\bibinfo{year}{2006}), \eprint{hep-ph/0509293}.

\bibitem[{\citenamefont{Zioutas et~al.}(1998)\citenamefont{Zioutas, Thompson,
  and Paschos}}]{Zioutas:1998ra}
\bibinfo{author}{\bibfnamefont{K.}~\bibnamefont{Zioutas}},
  \bibinfo{author}{\bibfnamefont{D.~J.} \bibnamefont{Thompson}},
  \bibnamefont{and} \bibinfo{author}{\bibfnamefont{E.~A.}
  \bibnamefont{Paschos}}, \bibinfo{journal}{Phys. Lett.}
  \textbf{\bibinfo{volume}{B443}}, \bibinfo{pages}{201} (\bibinfo{year}{1998}),
  \eprint{astro-ph/9808113}.

\bibitem[{\citenamefont{van Bibber et~al.}(1989)\citenamefont{van Bibber,
  McIntyre, Morris, and Raffelt}}]{vanBibber:1988ge}
\bibinfo{author}{\bibfnamefont{K.}~\bibnamefont{van Bibber}},
  \bibinfo{author}{\bibfnamefont{P.~M.} \bibnamefont{McIntyre}},
  \bibinfo{author}{\bibfnamefont{D.~E.} \bibnamefont{Morris}},
  \bibnamefont{and} \bibinfo{author}{\bibfnamefont{G.~G.}
  \bibnamefont{Raffelt}}, \bibinfo{journal}{Phys. Rev.}
  \textbf{\bibinfo{volume}{D39}}, \bibinfo{pages}{2089} (\bibinfo{year}{1989}).

\bibitem[{\citenamefont{Meeus}(1988)}]{Meeus}
\bibinfo{author}{\bibfnamefont{J.}~\bibnamefont{Meeus}},
  \emph{\bibinfo{title}{Astronomical Formul{\ae} for Calculators}}
  (\bibinfo{publisher}{Willmann-Bell}, \bibinfo{year}{1988}),
  \bibinfo{edition}{4th} ed.

\bibitem[{\citenamefont{F.~R.~Hoots}(1980)}]{spacetrack3}
\bibinfo{author}{\bibfnamefont{R.~L.~R.} \bibnamefont{F.~R.~Hoots}},
  \emph{\bibinfo{title}{Models for propagation of NORAD element sets}},
  \bibinfo{edition}{spacetrack report no. 3} ed. (\bibinfo{year}{1980}).

\bibitem[{tle()}]{tle}
\bibinfo{note}{{\tt http://celestrak.com}}.

\bibitem[{pre()}]{predict}
\bibinfo{note}{{\tt http://www.qsl.net/kd2bd/predict.html}}.

\bibitem[{\citenamefont{Kelso}(2007)}]{tleerror}
\bibinfo{author}{\bibfnamefont{T.~S.} \bibnamefont{Kelso}}, in
  \emph{\bibinfo{booktitle}{17th AAS/AIAA Space Flight Mechanics Conference}}
  (\bibinfo{year}{2007}), \bibinfo{number}{AAS 07-127}.

\bibitem[{wmm()}]{wmm}
\bibinfo{note}{{\tt http://www.ngdc.noaa.gov/seg/WMM/soft.shtml}}.

\bibitem[{\citenamefont{McLean et~al.}(2004)\citenamefont{McLean, Macmillan,
  Maus, Lesur, Thomson, and Dater}}]{wmm1}
\bibinfo{author}{\bibfnamefont{S.}~\bibnamefont{McLean}},
  \bibinfo{author}{\bibfnamefont{S.}~\bibnamefont{Macmillan}},
  \bibinfo{author}{\bibfnamefont{S.}~\bibnamefont{Maus}},
  \bibinfo{author}{\bibfnamefont{V.}~\bibnamefont{Lesur}},
  \bibinfo{author}{\bibfnamefont{A.}~\bibnamefont{Thomson}}, \bibnamefont{and}
  \bibinfo{author}{\bibfnamefont{D.}~\bibnamefont{Dater}}, \bibinfo{type}{Tech.
  Rep.} \bibinfo{number}{NOAA Technical Report NESDIS/NGDC-1},
  \bibinfo{institution}{NOAA National Geophysical Data Center}
  (\bibinfo{year}{2004}).

\bibitem[{\citenamefont{Coffey and Erwin}(2001)}]{coffey}
\bibinfo{author}{\bibfnamefont{H.~E.} \bibnamefont{Coffey}} \bibnamefont{and}
  \bibinfo{author}{\bibfnamefont{E.~H.} \bibnamefont{Erwin}},
  \bibinfo{journal}{Journal of Atmospheric and Solar-Terrestial Physics}
  \textbf{\bibinfo{volume}{63}}, \bibinfo{pages}{551} (\bibinfo{year}{2001}).

\bibitem[{\citenamefont{Siebert}(1971)}]{handbuchderphysik}
\bibinfo{author}{\bibfnamefont{M.}~\bibnamefont{Siebert}},
  \emph{\bibinfo{title}{Ma{\ss}zahlen der erdmagnetischen Aktivit\"at}}
  (\bibinfo{publisher}{Springer}, \bibinfo{year}{1971}),
  vol.~\bibinfo{volume}{49} of \emph{\bibinfo{series}{Handbuch der Physik}},
  pp. \bibinfo{pages}{206--275}.

\bibitem[{\citenamefont{Maus}()}]{maus}
\bibinfo{author}{\bibfnamefont{S.}~\bibnamefont{Maus}}, \bibinfo{note}{private
  communication}.

\bibitem[{\citenamefont{\protect{European cooperation for space
  standardization}}(2000)}]{ecss}
\bibinfo{author}{\bibnamefont{\protect{European cooperation for space
  standardization}}}, \bibinfo{type}{Tech. Rep.}
  \bibinfo{number}{ECSS-E-10-04A}, \bibinfo{institution}{European Space Agency}
  (\bibinfo{year}{2000}).

\bibitem[{\citenamefont{Henke et~al.}(1993)\citenamefont{Henke, Gullikson, and
  Davis}}]{henke}
\bibinfo{author}{\bibfnamefont{B.~L.} \bibnamefont{Henke}},
  \bibinfo{author}{\bibfnamefont{E.~M.} \bibnamefont{Gullikson}},
  \bibnamefont{and} \bibinfo{author}{\bibfnamefont{J.~C.} \bibnamefont{Davis}},
  \bibinfo{journal}{Atomic Data and Nuclear Data Tables}
  \textbf{\bibinfo{volume}{54}}, \bibinfo{pages}{181} (\bibinfo{year}{1993}).

\bibitem[{xra()}]{xraydata}
\bibinfo{note}{{\tt
  http://www.cxro.lbl.gov/optical$\underline~$constants/gastrn2.html}}.

\bibitem[{\citenamefont{Picone et~al.}(2002)\citenamefont{Picone, Hedin, Drob,
  and Aikin}}]{nrlmsise}
\bibinfo{author}{\bibfnamefont{J.~M.} \bibnamefont{Picone}},
  \bibinfo{author}{\bibfnamefont{A.~E.} \bibnamefont{Hedin}},
  \bibinfo{author}{\bibfnamefont{D.~P.} \bibnamefont{Drob}}, \bibnamefont{and}
  \bibinfo{author}{\bibfnamefont{A.~C.} \bibnamefont{Aikin}},
  \bibinfo{journal}{J. Geophys. Res.} \textbf{\bibinfo{volume}{107}},
  \bibinfo{pages}{1468} (\bibinfo{year}{2002}).

\bibitem[{\citenamefont{{Jacchia}}(1971)}]{Jacchia}
\bibinfo{author}{\bibfnamefont{L.~G.} \bibnamefont{{Jacchia}}},
  \bibinfo{journal}{SAO Special Report} \textbf{\bibinfo{volume}{332}}
  (\bibinfo{year}{1971}).

\bibitem[{\citenamefont{King-Hele and Hingston}(1967)}]{king}
\bibinfo{author}{\bibfnamefont{D.~G.} \bibnamefont{King-Hele}}
  \bibnamefont{and} \bibinfo{author}{\bibfnamefont{J.}~\bibnamefont{Hingston}},
  \bibinfo{journal}{Planet. Space Sci.} \textbf{\bibinfo{volume}{15}},
  \bibinfo{pages}{1883} (\bibinfo{year}{1967}).

\bibitem[{\citenamefont{Bowman}(1975)}]{bowman}
\bibinfo{author}{\bibfnamefont{B.~B.} \bibnamefont{Bowman}},
  \bibinfo{journal}{Planet. Space Sci.} \textbf{\bibinfo{volume}{23}},
  \bibinfo{pages}{1659} (\bibinfo{year}{1975}).

\bibitem[{\citenamefont{Hedin}(1991)}]{msise}
\bibinfo{author}{\bibfnamefont{A.~E.} \bibnamefont{Hedin}},
  \bibinfo{journal}{J. Geophys. Res.} \textbf{\bibinfo{volume}{96}},
  \bibinfo{pages}{1159} (\bibinfo{year}{1991}).

\bibitem[{\citenamefont{Raffelt}()}]{georg}
\bibinfo{author}{\bibfnamefont{G.}~\bibnamefont{Raffelt}},
  \bibinfo{note}{private communication}.

\bibitem[{\citenamefont{Carter and Read}(2007)}]{xmm}
\bibinfo{author}{\bibfnamefont{J.~A.} \bibnamefont{Carter}} \bibnamefont{and}
  \bibinfo{author}{\bibfnamefont{A.~M.} \bibnamefont{Read}}
  (\bibinfo{year}{2007}), \eprint{astro-ph/0701209}.

\bibitem[{\citenamefont{Mitsuda et~al.}(2007)}]{suzaku}
\bibinfo{author}{\bibfnamefont{K.}~\bibnamefont{Mitsuda}} \bibnamefont{et~al.},
  \bibinfo{journal}{Publ. Astron. Soc. Jap.} \textbf{\bibinfo{volume}{59}},
  \bibinfo{pages}{1} (\bibinfo{year}{2007}).

\bibitem[{\citenamefont{Revnivtsev et~al.}(2003)\citenamefont{Revnivtsev,
  Gilfanov, Sunyaev, Jahoda, and Markwardt}}]{xte}
\bibinfo{author}{\bibfnamefont{M.}~\bibnamefont{Revnivtsev}},
  \bibinfo{author}{\bibfnamefont{M.}~\bibnamefont{Gilfanov}},
  \bibinfo{author}{\bibfnamefont{R.}~\bibnamefont{Sunyaev}},
  \bibinfo{author}{\bibfnamefont{K.}~\bibnamefont{Jahoda}}, \bibnamefont{and}
  \bibinfo{author}{\bibfnamefont{C.}~\bibnamefont{Markwardt}},
  \bibinfo{journal}{Astron. Astrophys.} \textbf{\bibinfo{volume}{411}},
  \bibinfo{pages}{329} (\bibinfo{year}{2003}), \eprint{astro-ph/0306569}.

\bibitem[{\citenamefont{Koyama et~al.}(2007)}]{suzaku1}
\bibinfo{author}{\bibfnamefont{K.}~\bibnamefont{Koyama}} \bibnamefont{et~al.},
  \bibinfo{journal}{Publ. Astron. Soc. Jap.} \textbf{\bibinfo{volume}{59}},
  \bibinfo{pages}{23} (\bibinfo{year}{2007}).

\bibitem[{\citenamefont{Parmar and Turner}(2006)}]{xeus}
\bibinfo{author}{\bibfnamefont{A.~N.} \bibnamefont{Parmar}} \bibnamefont{and}
  \bibinfo{author}{\bibfnamefont{M.~J.~L.} \bibnamefont{Turner}},
  \bibinfo{type}{Tech. Rep.} \bibinfo{number}{SA/05.001/AP/cv},
  \bibinfo{institution}{ESA} (\bibinfo{year}{2006}).

\bibitem[{\citenamefont{Eidelman et~al.}(2004)}]{pdg}
\bibinfo{author}{\bibfnamefont{S.}~\bibnamefont{Eidelman}} \bibnamefont{et~al.}
  (\bibinfo{collaboration}{Particle Data Group}), \bibinfo{journal}{Phys.
  Lett.} \textbf{\bibinfo{volume}{B592}}, \bibinfo{pages}{1}
  (\bibinfo{year}{2004}).

\bibitem[{\citenamefont{Vallado}(2001)}]{vallado}
\bibinfo{author}{\bibfnamefont{D.}~\bibnamefont{Vallado}},
  \emph{\bibinfo{title}{Fundamentals of astrodynamic and applications}}
  (\bibinfo{publisher}{El Segundo California, Microcosm},
  \bibinfo{year}{2001}), \bibinfo{edition}{2nd} ed.

\bibitem[{wgs(2000)}]{wgs84}
\bibinfo{type}{Tech. Rep.} \bibinfo{number}{NIMA TR8350.2},
  \bibinfo{institution}{National Imagery and Mapping Agency}
  (\bibinfo{year}{2000}).

\end{thebibliography}

\newpage
\begin{appendix}

\section{Coordinate systems}
\label{app:coords}

Unfortunately,
the different algorithms and programs used for the calculation use
different coordinates systems which sometimes have non-trivial
transformation properties.  Therefore, a little digression on
commonly used coordinate systems is required. All coordinate systems
which can serve as quasi-inertial frames are ultimately defined by
astronomical observations. The idea is that very distant astronomical
objects like quasars allow us to define the orientation of a triad in
space which does not change with time. This triad then can be attached
to the barycenter of the Earth. We call this the celestial reference
system (CRS). Clearly, the CRS is not exactly inertial, but deviations
are very small of order $10^{-8}$ for special relativistic corrections
and $10^{-10}$ for general relativistic corrections~\cite{vallado}.
For example, the equations of motion of a satellite around the Earth,
or of the Earth around the Sun are valid in the CRS. Observers
typically do not float in space but are attached to the surface of the
Earth and thus observations will be relative to this surface, which
can be used to define the so called terrestial reference system (TRS),
which due to the rotation of the Earth is clearly not an inertial
system.

The task is to find the coordinate transformation from CRS to TRS and
its inverse, this process is also referred to as coordinate reduction.
This problem is, however, very much complicated by obsolete notations
stemming from times when the distinction between astrology and
astronomy was not always clear. Moreover, many layers of
approximations of varying accuracy are present, owing to the
difficulty of implementing a full coordinate reduction without
powerful enough computers.

Since both the CRS and TRS have the same origin in the barycenter of
the Earth\footnote{Note, that the International Celestial Reference
  Frame (ICRF) has its origin at the barycenter of the solar system.
  All coordinates we use have their origin at the barycenter of the
  Earth. The transformation between our Earth-centered CRF and the
  ICRF is a simple translation.}, the full transformation can be
described by three time-dependent rotations. In reality, these three
rotations are often split into four parts, since this makes it easier
to derive suitable approximations:

\begin{equation}
  \vec{x}_\mathrm{CRS}=\mathbf{P}(t)\mathbf{N}(t)
\mathbf{R}(t)\mathbf{W}(t)\vec{x}_\mathrm{TRS}\,,
\end{equation}
where $\mathbf{P}(t)$ accounts for the precession and $\mathbf{N}(t)$
for the nutation of the Earth spin axis. $\mathbf{R}(t)$ describes the
rotation of the Earth and $\mathbf{W}(t)$ the motion or wobble of the
spin axis with respect to the surface of the Earth.  The
precession is caused mainly by the Sun and the Moon pulling at the
equatorial bulge of the Earth, this is called the luni-solar
precession. Also, there is some precession of the ecliptic due to the
influence of the other planets on the Earth orbit around the Sun,
called planetary precession. The combined effects move the Earth axis
by about $50''$ per year. This effect was already known to the ancient
Greeks and was supposedly discovered by Hipparchos. Nutation is
mainly caused by the fact that the Moon's orbit is inclined with
respect to the Earth's equator and hence the Moon's pull on the
equatorial bulges changes throughout each month. The full theory
of nutation is quite complicated since it receives contributions from
many sources. The result is as a main period of 18.6 years and an
amplitude varying from $9''$ to $17''$.

The Earth rotation is not uniform either and changes {\it e.g.} due to
the friction caused by the tidal bulges. In the simplest case the
required rotation angle would be proportional to
\begin{equation}
\label{eq:ut1}
\omega_{\oplus} t\,.
\end{equation}
Instead of defining a time dependent angular velocity, all violations
of this simple relation are absorbed into the definition of time. The
relevant time system is UT1 (Universal Time 1) which is based on
observed transit times of distant astronomical objects and basically
ensures the validity of Eq.~\ref{eq:ut1}. UTC (Universal Time
Coordinated) is the one on which our daily life is based on. It is
adjusted to keep track of UT1 by insertion of leap seconds whenever
required and never differs by more than $\pm0.9\,\mathrm{s}$ from
UT1.

The wobble of the spin axis exists since the Earth is not a rigid
body, but has a liquid interior. It is a very difficult effect to predict;
fortunately it is a very small effect of only $0.1''$.

In all coordinate transformations used throughout this work, we will
neglect effects caused by nutation, the difference between UT1 and UTC
as well as any polar wobble, therefore our basic transformation
reduces to\footnote{This choice, in particular, implies that we do not
  convert TEME as used by SGP4 into a Mean of Date system.}
$\vec{x}_\mathrm{CRS}=\mathbf{P}(t)\mathbf{R}(t)\vec{x}_\mathrm{TRS}$.
This choice of approximations is mainly guided by the obtainable
accuracy of the satellite orbit prediction system used. The errors
induced are about $\pm1\,\mathrm{s}$ in timing and less than $30''$ in
angle.

Next we need to define the Terrestrial Reference System (TRS) or
geodetic coordinates (GC). Positions on the Earth are commonly
measured by latitude, longitude and height above mean sea level. The
Earth is not spherically symmetric, but to a very good
approximation\footnote{In reality, the shape of the Earth, the geoid,
  is defined as being an equipotential surface of its gravitational
  potential. There is no simple closed analytic form for the geoid.
  The deviations from the ellipsoid are called undulation of the geoid
  and are indeed very small $<200\,\mathrm{m}$. Note, that mountains
  which can be up to $\sim10\,000\,\mathrm{m}$ do not play a role in
  this paper since x-ray propagation ceases at altitudes well above
  that, thus only the shape of the geopotential iso-surfaces, which
  determine air density, at heights above about $50\,\mathrm{km}$ are
  relevant.} an oblate ellipsoid, with a flattening of about $1/300$.
This is caused by the centrifugal force due to the Earth rotation and
the fact the Earth is not a rigid but elastic body.  There are two
consequences from this definition of GC: the vector normal to the
Earth surface no longer points back to the center of the Earth and the
latitude is now defined as the angle between the normal to the surface
and the equator. The satellite propagation routines use the so called
World Geodetic System 72 (WGS72), whereas the geomagnetic model is
based on WGS84. The difference between these two system is less than
$6''$ in latitude, less than $1''$ in longitude and less than
$6\,\mathrm{m}$ in height~\cite{wgs84}. Therefore we use them
interchangeably in coordinate transformations, whereas the satellite
propagation routines use WGS72.

With the exception of GC, all other
coordinates system are just simple Cartesian or polar coordinate
systems which are related by standard transformations. All numerical
algorithms follow the description in chapter~3 of~\cite{vallado}. We
will therefore give the correspondence of our notation with the one
of~\cite{vallado} in Table~\ref{tab:coords}.
\begingroup
\squeezetable

\begin{longtable}{|cccccc|}
  \hline
  &&Fundamental&Principal&&\\
  Symbol&Origin&Plane&direction&Use&Notation in~\cite{vallado}\\
  \hline
  ECI&Earth&Earth equator&Vernal equinox&Main system for numerical calculations&IJK\\
  TRS&Earth&Earth equator&Greenwich meridian&Intermediate step in
  coordinate conversion&$(\mathrm{IJK})_{\mathrm{ITRF}}$\\
  GC&Earth&Earth equator&Greenwich meridian&Input to geomagnetic
  model&LatLon\\
  TCM&Site&Local horizon&North&Output of geomagnetic model&-\\
  TC&Site&Local horizon&South&Intermediate step in coordinate conversion&SEZ\\
\hline

\caption{\label{tab:coords} Coordinate systems used and their
  corresponding names used in~\cite{vallado}. Note that some
  coordinate systems used are Cartesian whereas others are polar.}
\end{longtable}
\endgroup

We will use Cartesian Earth centered inertial coordinates (ECI) for
all numerical calculations and thus convert all coordinates into ECI
first. The motion of the satellite is directly evaluated in ECI. Also
the position of the Sun is directly given in ECI. The Earth magnetic
field is specified by its location in geodetic coordinates (GC) and
the result is a vector in topocentric coordinates (TCM). The density of
the Earth atmosphere is a function of the altitude which is defined in
GC. Together with Table~\ref{tab:coords} we obtain the following chain of
coordinate transformations for the magnetic field $\vec{B}$:

\begin{equation}
\vec{x}_\mathrm{ECI}\longrightarrow\vec{x}_\mathrm{TRS}\longrightarrow\vec{B}_\mathrm{TCM}=\vec{B}(\vec{x}_\mathrm{GC})\stackrel{\mathbf{R}_3(\pi)}{\longrightarrow}\vec{B}_\mathrm{TC}\longrightarrow\vec{B}_\mathrm{ECI}\,,
\end{equation}
where only the non-standard transformations are given on top of each
arrow, which is a rotation around the 3 or z-axis to get from TCM to
TC.

\section{Signal extraction region and angular resolution}
\label{app:signal}

\begin{figure}[t!]
\includegraphics[width=0.5\textwidth]{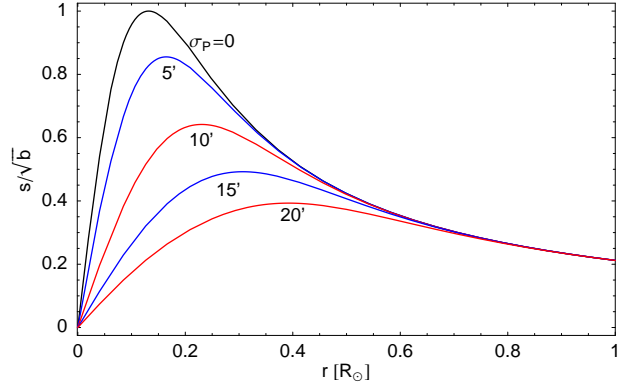}
\caption{\label{fig:psf} 
Relative significance $s/\sqrt{b}$ as a function of the solar
radius. The different curves are for different values of the width of
the point spread function as labeled in the plot.
}
\end{figure}

The Sun has an average angular diameter of about $32'$ and the axion
producing region is mostly confined to the inner 20-30\% of the solar
radius. The typical angular resolution of x-ray telescopes ranges from
arc seconds to a few arc minutes, therefore the Sun is not a point
source of axions. As a result, the ``night-side" image of the Sun in
x-rays from GECOSAX will cover a finite area in the focal plane of the
telescope and it is possible to select a spot radius $r$ which
optimizes the significance $s/\sqrt{b}$. Both the signal and the
background will be a function of $r$, which we take be a dimensionless
fraction of the solar radius. Since axion production is very much
concentrated towards the center of the Sun, $s$ steeply rises for
small values of $r$ and then saturates at $r\simeq0.3$, whereas the
background, assuming it is spatially uniform, will rise as $r^2$. Thus
there should be a maximum in the significance and the radius for which
this happens is called $r_s$. Assuming perfect spatial resolution of
the telescope and no pointing errors, we find $r_s=0.13$. In the
following we will rescale all values of $s/\sqrt{b}$ by the value
obtained at $r_s$ in this case. The rescaled $s/\sqrt{b}$ as a
function of $r$ is shown as the black line in Fig.~\ref{fig:psf}.

So far we have assumed that the telescope has perfect resolution and
that there are no pointing errors. Both errors have the effect
that they will blur the image of the Sun and thus the signal
density per unit area will decrease, whereas the background density is
unaffected. Therefore the significance will decrease relative to the
ideal case and at the same time $r_s$, the optimal signal extraction
radius will increase. We assume the point spread function, which
contains both the effects from pointing errors and the finite optical
resolution of the telescope to be a Gaussian with a width or standard
deviation of $\sigma_p$. The result of finite values for $\sigma_p$
are shown as blue and red lines in Fig.~\ref{fig:psf}. 
The values of $\sigma_p$ 
next to each line are in arc minutes. The resulting
value at maximum is the correction factor which needs to be applied to
the product $\Sigma Q$ in Eq.~(\ref{eq:limit}) in order to estimate
the limiting sensitivity to $g_{a\gamma}$. The SUZAKU telescope has a
resolution of better than $2.5'$ and pointing accuracy of better than
$0.25'$~\cite{suzaku}, thus the resulting correction factor is $0.96$,
which was neglected in computing figure~\ref{fig:sens}.

\newpage
\section{Satellite TLE$\mathrm{s}$ and orbit parameters}

\begingroup
\squeezetable

\begin{longtable}{||cl|rrr|rrrrr|rrrrr|c||}

  \caption{\label{tab:tle} This table lists all satellites used
    throughout this study. Given are the US SPACECOM identification
    numbers, the name of the satellite and its general orbit
    parameters. The fixed and turning mode columns show $80\%$
    significance $\Sigma_{80}$ as defined in Eq.~\ref{eq:sigma}, the
    required quality factor to achieve a sensitivity as good as the
    CAST experiment using $\Sigma_{80}$, the number of dark orbits
    $n_{80}$ needed to reach $\Sigma_{80}$, and the total time needed to
    achieve $\Sigma_{80}$ (this includes the contribution of
    $t_\mathrm{cut}$ for each orbit). The last column for each mode is
    the corresponding rank within the satellites in this table. '-'
    indicates that there is no useful dark orbit left after cutting away
    $600\,\mathrm{s}$. TLE set is the number of the TLE set used for
    the calculations in this paper. All satellite data and TLEs are
    from~\cite{tle}.}\\

\hline
\hline
ID&Name&Perigee&Apogee&Inclination&\multicolumn{5}{c|}{fixed
  mode}&\multicolumn{5}{c|}{turning mode}&TLE set\\
&&&&&$\Sigma_{80}$&$Q_\mathrm{CAST}$&$n_{80}$&$t_{80}$&rank&$\Sigma_{80}$&$Q_\mathrm{CAST}$&$n_{80}$&$t_{80}$
&rank&\\
&&km&km&$\circ$&\rotatebox{90}{$10^{-3}\,\mathrm{cm}^{-2}\,\mathrm{s}^{-1/2}$}&\rotatebox{90}{$10^3\,\mathrm{cm}^{2}\,\mathrm{s}^{1/2}$}&
&\rotatebox{90}{$10^{6}\,\mathrm{s}$}&&\rotatebox{90}{$10^{-3}\,\mathrm{cm}^{-2}\,\mathrm{s}^{-1/2}$}&\rotatebox{90}{$10^3\,\mathrm{cm}^{2}\,\mathrm{s}^{1/2}$}& &$\rotatebox{90}{$10^{6}\,\mathrm{s}$}$&&\\
\hline
\hline
\endfirsthead

\caption{continued}\\
\hline
\hline
ID&Name&Perigee&Apogee&Inclination&\multicolumn{5}{c|}{fixed
  mode}&\multicolumn{5}{c|}{turning mode}&TLE set\\
&&&&&$\Sigma_{80}$&$Q_\mathrm{CAST}$&$n_{80}$&$t_{80}$&rank&$\Sigma_{80}$&$Q_\mathrm{CAST}$&$n_{80}$&$t_{80}$
&rank&\\
&&km&km&$\circ$&\rotatebox{90}{$10^{-3}\,\mathrm{cm}^{-2}\,\mathrm{s}^{-1/2}$}&\rotatebox{90}{$10^3\,\mathrm{cm}^{2}\,\mathrm{s}^{1/2}$}&
&\rotatebox{90}{$10^{6}\,\mathrm{s}$}&&\rotatebox{90}{$10^{-3}\,\mathrm{cm}^{-2}\,\mathrm{s}^{-1/2}$}&\rotatebox{90}{$10^3\,\mathrm{cm}^{2}\,\mathrm{s}^{1/2}$}& &$\rotatebox{90}{$10^{6}\,\mathrm{s}$}$&&\\
\hline
\hline
\endhead
\hline
\multicolumn{16}{||c||}{continued on next page}\\
\hline
\hline
\endfoot

\endlastfoot

13777&IRAS&888&924&80.9&2.8&1.2&588&0.5&2&-&-&-&-&-&404\\
20322&COBE&874&909&81.1&2.5&1.3&445&0.4&4&-&-&-&-&-&548\\
20580&HST&561&574&28.6&0.7&5.0&2301&4.6&45&0.3&13.2&1931&3.9&35&70\\
20638&ROSAT&398&419&53.1&0.6&5.9&1754&3.1&48&0.2&17.1&1170&2.2&38&293\\
21578&SARA&725&754&81.9&2.5&1.3&2061&2.9&5&0.7&4.8&748&1.2&3&148\\
21701&UARS&368&475&57.1&0.7&4.8&1441&2.4&44&0.2&14.4&1067&2.0&37&290\\
22012&SAMPEX&427&493&81.6&0.9&3.6&1472&2.2&40&0.3&11.9&820&1.4&33&525\\
23547&ORBVIEW 1 (MICROLAB)&705&728&70.1&1.7&1.9&1463&2.5&20&0.6&5.3&783&1.4&8&347\\
23757&XTE&479&492&23.1&0.4&7.5&2420&5.0&50&0.2&20.2&2090&4.3&40&549\\
25280&TRACE&567&600&82.2&1.9&1.8&1418&1.5&17&0.5&7.2&807&1.2&17&753\\
25399&SAFIR 2&814&839&81.6&2.8&1.2&1950&2.7&1&0.8&4.2&717&1.2&1&963\\
25560&SWAS&605&628&70.0&1.7&1.9&1008&1.3&21&0.5&7.0&785&1.3&14&577\\
25635&ORSTED&652&869&83.5&1.7&2.0&2248&4.6&24&0.5&6.3&1709&3.5&10&287\\
25636&SUNSAT&653&879&83.4&1.7&1.9&2205&4.4&19&0.5&6.1&1762&3.5&9&934\\
25646&WIRE&412&434&82.7&0.6&5.8&2459&5.2&47&0.2&21.4&2083&4.4&41&414\\
25721&ABRIXAS&504&526&48.6&0.9&3.9&1828&3.2&41&0.3&11.1&1261&2.4&30&221\\
25735&TERRIERS&493&526&82.7&1.0&3.5&2448&4.9&39&0.3&11.5&2070&4.1&31&354\\
25791&FUSE&743&765&25.1&1.0&3.2&2254&4.5&37&0.4&8.2&1952&3.9&24&116\\
25978&CLEMENTINE&610&647&81.8&1.4&2.4&2201&4.2&32&0.4&7.8&1714&3.3&23&271\\
25994&TERRA&704&730&81.8&1.5&2.2&2307&4.7&29&0.4&7.5&1854&3.8&20&308\\
26033&ACRIMSAT&680&739&81.9&1.5&2.2&2360&4.8&28&0.5&7.1&2003&4.1&16&309\\
26546&MEGSAT-1&607&639&64.7&1.5&2.3&1384&2.3&30&0.5&6.9&884&1.6&13&316\\
26561&HETE-2&560&594&2.0&0.5&6.6&2504&5.3&49&0.2&17.2&2219&4.7&39&231\\
26702&ODIN&582&611&82.1&1.7&2.0&614&0.7&23&0.3&10.1&285&0.4&28&310\\
26998&TIMED&619&643&74.1&1.3&2.5&1552&2.6&33&0.5&7.1&764&1.4&15&943\\
27370&RHESSI&551&583&38.2&0.8&4.0&2049&3.8&42&0.3&11.0&1638&3.2&29&723\\
27598&FEDSAT&795&827&81.6&2.1&1.6&2179&4.1&12&0.7&4.9&1687&3.2&4&418\\
27599&WEOS&792&827&81.5&2.0&1.7&2204&4.3&16&0.6&5.3&1731&3.4&7&436\\
27600&MICRO LABSAT&791&827&81.5&2.0&1.7&2206&4.3&15&0.6&5.3&1730&3.4&6&411\\
27640&CORIOLIS&823&866&81.2&2.3&1.4&507&0.4&6&-&-&-&-&-&313\\
27643&CHIPSAT&573&598&86.0&1.2&2.7&1359&2.2&34&0.4&9.2&903&1.7&27&685\\
27651&SORCE&612&642&40.1&1.1&3.1&1942&3.5&35&0.4&8.6&1546&2.9&25&531\\
27783&GALEX&692&698&29.1&1.0&3.4&2217&4.3&38&0.4&8.8&1879&3.7&26&331\\
27843&MOST&825&847&81.2&2.3&1.4&521&0.4&7&-&-&-&-&-&235\\
27845&QUAKESAT&826&847&81.2&2.3&1.4&522&0.4&9&-&-&-&-&-&215\\
27846&AAU CUBESAT&820&846&81.3&2.3&1.4&525&0.4&8&-&-&-&-&-&163\\
27858&SCISAT 1&643&670&73.9&2.2&1.5&961&1.1&10&0.5&6.7&913&1.5&12&379\\
27945&KAISTSAT&677&715&81.9&1.6&2.1&2364&4.6&25&0.5&6.5&1993&3.9&11&347\\
28230&GP-B&644&669&87.8&1.8&1.8&1266&1.6&18&0.4&7.8&973&1.6&22&147\\
28368&DEMETER&664&690&81.9&1.4&2.3&2301&4.6&31&0.4&7.7&1822&3.7&21&185\\
28485&SWIFT&583&598&20.7&0.6&5.3&2390&4.9&46&0.2&14.0&2092&4.3&36&935\\
28773&SUZAKU&559&576&31.5&0.7&4.7&2269&4.5&43&0.3&12.3&1893&3.8&34&675\\
28939&ASTRO-F (AKARI)&699&733&81.7&2.0&1.6&571&0.5&13&-&-&-&-&-&628\\
29052&FORMOSAT 3&777&838&72.1&2.1&1.6&1322&2.1&11&0.7&4.5&757&1.3&2&651\\
29107&CLOUDSAT&704&730&81.8&1.5&2.2&2417&4.9&27&0.5&7.2&1885&3.8&19&602\\
29108&CALIPSO&704&730&81.8&1.5&2.2&2417&4.9&26&0.5&7.2&1885&3.8&18&587\\
29479&HINODE (SOLAR-B)&687&709&81.9&2.0&1.7&635&0.6&14&-&-&-&-&-&462\\
29506&SJ-6D&600&628&82.3&1.7&2.0&685&0.7&22&0.1&26.9&127&0.2&42&495\\
29678&COROT&901&930&89.2&2.6&1.3&1287&1.8&3&0.6&5.2&1032&1.8&5&376\\
31304&AIM&586&620&81.8&1.0&3.2&2403&5.1&36&0.3&11.5&2005&4.3&32&275\\
\hline
\hline

25544&ISS (ZARYA)\footnote{ The ISS orbit is subject to frequent
  changes due to maneuvering, therefore this result can
  only serve as a rough indicator of the ISS' orbit quality.}&347&365&51.8&0.4&8.9&1741&3.1&-&0.1&26.4&1158&2.2&-&125\\

21118&MOLNIYA 1-80\footnotemark[28]&792&39608&63.5&1.9&1.8&79&0.2&-&0.9&3.6&73&0.2&-&434\\
21196&MOLNIYA 3-40\footnotemark[28]&721&39664&63.2&1.9&1.8&87&0.2&-&1.0&3.5&84&0.2&-&733\\
22729&MOLNIYA 3-45\footnotemark[28]&683&39725&63.8&2.0&1.7&42&0.1&-&0.9&3.8&49&0.1&-&589\\
23211&MOLNIYA
3-46\footnotemark[28]&517&39776&62.2&2.0&1.7&31&0.1&-&0.8&4.3&27&0.1&-&347\\

\hline
\hline

\end{longtable}

\endgroup
\footnotetext[28]{These orbits exceed the validity range of the
  geomagnetic model and maybe subject to increased uncertainties. To
  minimize this effect, only those parts of the orbit with altitude
  below $1R_\oplus$ are considered. }

\end{appendix}

\end{document}